\documentclass[namedreferences]{solarphysics}
\usepackage[hyperref,optionalrh,solaromanenum]{spr-sola-addons} 
\usepackage{graphicx}                    
\usepackage{amssymb}                   
\usepackage{color}                          
\usepackage{breakurl}                     

\usepackage{amsmath}
\usepackage{natbib}
\usepackage{color}
\usepackage{placeins}
\usepackage{rotating}

\newcommand{\ie}{{\it i.e. }}
\newcommand{\eg}{{\it e.g. }}
\newcommand{\insitu}{{\it in situ }}
\newcommand{\Rsun}{{$~\mathrm{R}_{\odot}$}}
\newcommand{\kmps}{{$\mathrm{~km~s}^{-1}$}}

\newcommand{\tbn}{{the initial magnetic field was estimated based on the GCS reconstruction and dimming flux (for details see main text)}}
\newcommand{\tbnn}{{for FDmax: value at EPHIN/value at SoPo}}
\newcommand{\tbnnn}{{flow speed at MES and VEX was estimated based on the method described by Vrsnak et al. (2019), assuming that the flow speed remains roughly constant beyond Earth distance}}
\newcommand{\tbnnnn}{{estimated based on Wind observations assuming constant expansion speed}}
\newcommand{\tbnnnnn}{{estimated based on the standoff distance between GCS reconstructed FR and shock}}
\newcommand{\tbnnnnnn}{{calculated for perfect detector which responds to all energies above cutoff equally}}
\newcommand{\tbnnnnnnn}{{calculated at Earth for EPHIN based on the response function}}

\begin{document}
\begin{article}
\begin{opening}
\title{Evolution of coronal mass ejections and the corresponding Forbush decreases: modelling \textit{vs} multi-spacecraft observations}
 \author[addressref={aff1},corref,email={mateja.dumbovic@geof.unizg.hr}]{\inits{M.}\fnm{Mateja}~\lnm{Dumbovi\' c}\orcid{0000-0002-8680-8267}}
 \author[addressref=aff1]{\inits{B.}\fnm{Bojan}~\lnm{Vr\v snak}\orcid{0000-0002-0248-4681}}
 \author[addressref={aff2,aff3}]{\inits{J.}\fnm{Jingnan}~\lnm{Guo}\orcid{0000-0002-8707-076X}}
 \author[addressref=aff4]{\inits{B.}\fnm{Bernd}~\lnm{Heber}\orcid{0000-0003-0960-5658}}
 \author[addressref=aff5]{\inits{K.}\fnm{Karin}~\lnm{Dissauer}\orcid{0000-0001-5661-9759}}
 \author[addressref=aff6]{\inits{F.}\fnm{Fernando}~\lnm{Carcaboso}\orcid{0000-0003-1758-6194}}
 \author[addressref=aff5]{\inits{M.}\fnm{Manuela}~\lnm{Temmer}\orcid{0000-0003-4867-7558}}
 \author[addressref={aff5,aff7}]{\inits{A.}\fnm{Astrid}~\lnm{Veronig}\orcid{0000-0003-2073-002X}}
 \author[addressref=aff8]{\inits{T.}\fnm{Tatiana}~\lnm{Podladchikova}\orcid{0000-0002-9189-1579}}
 \author[addressref=aff9]{\inits{C.}\fnm{Christian}~\lnm{M\"ostl}\orcid{0000-0001-6868-4152}}
 \author[addressref=aff9]{\inits{T.}\fnm{Tanja}~\lnm{Amerstorfer}\orcid{0000-0001-9024-6706}}
 \author[addressref=aff10]{\inits{A.}\fnm{Anamarija}~\lnm{Kirin}\orcid{0000-0002-6873-5076}}
 \address[id=aff1]{Hvar Observatory, Faculty of Geodesy, University of Zagreb, Ka\v{c}i\'ceva 26, HR-10000 Zagreb, Croatia}
 \address[id=aff2]{CAS Key Laboratory of Geospace Environment, School of Earth and Space Sciences, University of Science and Technology of China, Hefei, China}
  \address[id=aff3]{CAS Center for Excellence in Comparative Planetology, USTC, Hefei, 230026, China}
 \address[id=aff4]{Department of Extraterrestrial Physics, Christian-Albrechts University in Kiel, Liebnitzstrasse 11, 24098, Kiel, Germany}
 \address[id=aff5]{Institute of Physics, University of Graz, Universit\"atsplatz 5, A-8010 Graz, Austria}
 \address[id=aff6]{Dpto. de Fisica y Matematicas, Universidad de Alcala, 28805 Alcala de Henares, Madrid, Spain}
 \address[id=aff7]{Kanzelh\"ohe Observatory for Solar and Environmental Research, University of Graz, Kanzelh\"ohe 19, 9521 Treffen, Austria}
 \address[id=aff8]{Skolkovo Institute of Science and Technology, Bolshoy Boulevard 30, bld. 1, Moscow 121205, Russia}
 \address[id=aff9]{Space Research Institute, Austrian Academy of Sciences, Schmiedlstra\ss e 6, 8042 Graz, Austria}
 \address[id=aff10]{Karlovac University of Applied Sciences, Karlovac, Croatia}
\runningauthor{Dumbovi\' c et al.}
\runningtitle{CME evolution \& FD}
\begin{abstract}
One of the very common \insitu signatures of interplanetary coronal mass ejections (ICMEs), as well as other interplanetary transients, are Forbush decreases (FDs), \ie short-term reductions in the galactic cosmic ray (GCR) flux. A two-step FD is often regarded as a textbook example, which presumably owes its specific morphology to the fact that the measuring instrument passed through the ICME head-on, encountering first the shock front (if developed), then the sheath and finally the CME magnetic structure. The interaction of GCRs and the shock/sheath region, as well as the CME magnetic structure, occurs all the way from Sun to Earth, therefore, FDs are expected to reflect the evolutionary properties of CMEs and their sheaths. We apply modelling to different ICME regions in order to obtain a generic two-step FD profile, which qualitatively agrees with our current observation-based understanding of FDs. We next adapt the models for energy dependence to enable comparison with different GCR measurement instruments (as they measure in different particle energy ranges). We test these modelling efforts against a set of multi-spacecraft observations of the same event, using the Forbush decrease model for the expanding flux rope (\textit{ForbMod}). We find a reasonable agreement of the \textit{ForbMod} model for the GCR depression in the CME magnetic structure with multi-spacecraft measurements, indicating that modelled FDs reflect well the CME evolution.
\end{abstract}
\keywords{Coronal Mass Ejections, Interplanetary; Cosmic Rays, Galactic}
\end{opening}
 \section{Introduction}
 \label{intro} 

Coronal mass ejections (CMEs) are magnetic structures that erupt from the solar corona and interact with the ambient plasma and energetic particles (electrons, protons and ions ranging from suprathermal to GeV energies and beyond), as they evolve and propagate through the interplanetary (IP) space. The propagation of CMEs in the IP space is dominated by the emission of MHD waves in the collisionless solar wind environment, that acts to adjust the CME speed to the ambient solar wind, \ie MHD drag \citep{cargill96, vrsnak13}. As they propagate through the IP space, CMEs expand due to the pressure imbalance \citep[\eg][and references therein]{klein82,demoulin09}, where, consequently, as the size of the magnetic structure increases, its magnetic field weakens \citep[\eg][]{bothmer98,leitner07,demoulin08,gulisano12,vrsnak19}. CMEs with leading fronts moving faster than the ambient solar wind will compress and deflect the upstream plasma producing a so-called sheath region, and if their relative speed is greater than the fast-mode wave speed, a shock will form \citep[\eg][]{russell02, owens05}. The sheath plasma can be composed of different types of material (coronal/heliospheric, shocked/compressed) and its size can also change, as the CME evolves and propagates \citep[][]{masias-meza16,janvier19,lugaz20}. 

Interplanetary coronal mass ejections (ICMEs), according to the standard nomenclature, encompass both a shock/sheath region and a CME magnetic structure \citep[\eg][]{rouillard11,kilpua17}, each showing a number of specific \insitu properties. The shock/sheath region is typically characterised by increased density, temperature, magnetic field fluctuations and plasma beta, whereas the CME magnetic structure is typically characterised by a smoothly rotating field, low plasma beta and temperature and a linearly decreasing speed profile indicative of expansion \citep[][]{zurbuchen06, kilpua17}. These properties indicate magnetic structures with field lines winding helicoidally around the central axis, therefore, the most commonly assumed simple magnetic structure of a CME is a force-free flux rope with circular cross section \citep[\eg][]{lundquist51,gold60} that expands self-similarly \citep[][]{demoulin08}. It should be noted however, that observed magnetic structures can substantially deviate from this highly ideal concept \citep[\eg][]{nieves-chinchilla18}, and might not even be flux ropes, but simply writhed structures \citep[][]{al-haddad19}. Additional indication of the CME magnetic structure can be increased abundance of high charge states, reflecting the temperature history of the CME and/or its origin \citep[][]{lepri01,zurbuchen16}, or bi-directional suprathermal (60-1000eV) electrons, indicative of a magnetically closed structure \citep[][]{gosling87}. ICMEs can also show distinctive signatures in galactic cosmic ray (GCR) count rates, in detectors which have count rates high enough to provide sufficient statistical accuracy, depressions (Forbush decreases, FDs) can be observed.

FDs can be observed throughout the heliosphere \citep[\eg][]{paularena01,witasse17,winslow18} using detectors such as ground-based neutron monitors and  muon detectors, spacecraft particle detectors and dosimeters and have recently been substantially utilised as ICME signatures at Mars \citep[\eg][]{forstner18,guo18b, forstner19, papaioannou19}. FD properties such as the magnitude, shape, duration and sub-structuring depend on the properties of the corresponding interplanetary transient \citep[see \eg][]{richardson04,cane00,belov09,dumbovic12b}. FDs caused by ICMEs often show a two-step profile, one associated with the shock/sheath region and the other with the CME magnetic structure \citep{barnden73, cane00}. The two regions were found to be roughly equally effective in producing the depression \citep{richardson11a}, although because of their different physical properties, the mechanism through which they produce the depression is different and thus should be modelled differently \citep{wibberenz98}. The shock acts as a discontinuity which can reflect particles \citep{kirin20}, resulting in a pre-increase \citep{cane00}, whereas the sheath can be described as a diffusive barrier \citep{wibberenz97,wibberenz98} and the shock/sheath-related FD can be described by solving a full transport \citep[][]{parker65} equation \citep[\eg][]{leroux91,wawrzynczak10,alania13}. On the other hand, a CME magnetic structure can be described as a very slowly filling (and expanding) particle trap where the particles can enter \textit{via} perpendicular diffusion \citep[\eg][]{cane95,munakata06,subramanian09,dumbovic18b}, or guiding center drifts \citep[\eg][]{krittinatham09,tortermpun18}. Recent efforts in FD modelling also include full trajectory integration using CME flux-rope type models \citep{petukhova19a,petukhova19b} or CME magnetic field reconstructions from \insitu measurements \citep{benella19}, as well as describing FDs \textit{via} the change in the single GCR spectrum modulation parameter attributed with a CME \citep{guo20}.

Recently, \citet{dumbovic19} utilised FD signatures at Mars to indicate inhibited expansion of a CME magnetic structure and \citet{forstner20} related different FD properties at Earth and Mars to the possible evolution of the ICME sheath. Therefore, there are strong indications that FDs reflect the evolutionary properties of ICMEs and thus FD models not only offer an opportunity to understand the variability of FDs detected in the heliosphere, but also to gain insight into ICME evolution. For that purpose we consider a text-book example of a two-step FD and combine two analytical models, the propagative diffusive barrier \citep[PDB,][]{wibberenz98} model for the sheath region and the diffusion-expansion Forbush decrease \citep[\textit{ForbMod},][]{dumbovic18b} model for the CME magnetic structure to produce a generic two-step FD profile (Section \ref{profile}). We adapt these models for energy dependence (Section \ref{energy}), since in their current form they are monoenergetic, whereas detectors which have enough statistics to detect FDs detect an energy range (usually $E>E_{\mathrm{cutoff}}$). Finally, using modelling, we analyse an actual multi-spacecraft event recently studied from the observational perspective by \citet{winslow18} at 4 different radially aligned heliospheric distances (Section \ref{obs}).

\section{A generic two-step Forbush decrease profile}
\label{profile}

A two-step Forbush decrease (FD) is often regarded as a textbook FD, which owes its specific morphology to the fact that the measuring instrument passed through the ICME head-on \citep[see e.g.][and references therein]{richardson11a}. In this case, the instrument first encounters the shock front (if developed), then the sheath and finally the CME magnetic structure. The corresponding depression has a sharp onset, sometimes preceded by a small pre-increase, where this first drop is then suddenly interrupted by the second onset, \ie the second decrease. The first decrease is usually attributed to the joint shock/sheath region, and the second decrease to the magnetic ejecta. The GCR count in a two-step FD does not return to the pre-decrease level immediately after the ICME passage, but recovers slowly over the course of the next couple of days or even weeks. In order to understand the mechanisms which govern the formation of the depression, we need to analyse the interplanetary structures that cause them and how they influence the GCRs.

 \begin{figure} 
 \centerline{\includegraphics[width=0.99\textwidth]{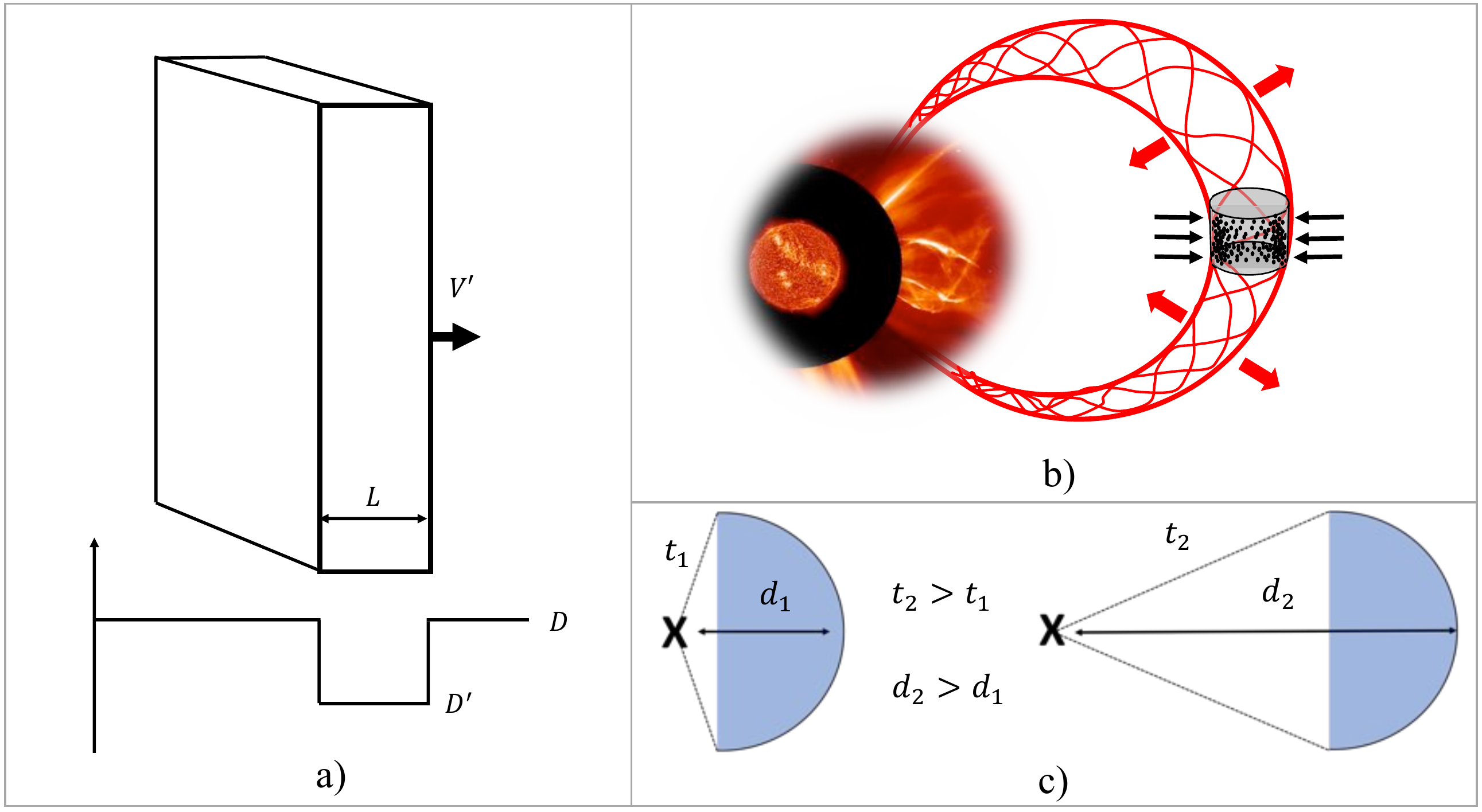}}
 \caption{Sketch of the models used for the generic profile of FD: a) propagating diffusive barrier \citep[PDB,][]{wibberenz98}, the sheath is represented by a shell with constant flow speed, $V'$ and decreased constant diffusion coefficient, $D'$; b) \textit{ForbMod} \citep{dumbovic18b}, GCRs (black arrow) diffuse into the expanding flux rope (red arrows); c) shadow effect of the shock \citep{lockwood86}, shock 'casts' larger shadow at the observer at time $t_1$ when it is closer to the observer.}
 \label{fig1}
\end{figure}

\subsection{The shock-related effect}
\label{shock}

Since the magnetic field is compressed in the downstream region of the interplanetary shock it can be regarded as a fast mode MHD shock. It was shown by \citet{kirin20} that due to the change of the magnetic field component normal to the shock front, particles coming from the upstream region can be reflected. This can cause or contribute to the first decrease in the two-step FD, but also explain the pre-increase observed in some events,  namely, the interaction of the GCR with a shock is a result of the complex interplay of the shock orientation and strength vs GCR energy and direction. Test particle simulations show that a subset of the GCR particles may be reflected between the upstream and downstream regions and reenter the shock region \textit{via} helical motion \citep[for details see][]{kirin20}. Therefore, a small population of particles will linger for some time near the shock front. Although the 'shock' effect is produced in the extremely small spatial extent of the shock front \citep[typically $\sim10^3$km;][]{pinter80} due to reflection of particles, it has influence over a much broader spatial scale. Very strong shocks are often associated with so-called precursors, pre-decreases and pre-increases in the CR intensity accompanied by the changes in the first harmonic of the anisotropy at the ecliptic plane, appearing hours before the FD onset \citep[see \eg][and references therein]{belov95,papailiou12,lingri19}.

Moreover, there are strong indications that a prolonged recovery phase of the FD is due to the shock front moving away from the observer. As it propagates away the shock reflects the upstream particles across its front, which can be regarded as ``casting a shadow" upon the observer. As the shock front propagates away the ``shadow" becomes smaller, thus the shadow effect weakens (i.e. decays) resulting in an exponential recovery \citep[for more details see][]{lockwood86,dumbovic11}. The shadow effect is sketched in Figure \ref{fig1}c. In several studies the recovery phase of FDs was successfully fitted by an exponential decay function \citep[\eg][]{penna05,jamsen07,usoskin08,zhao16b,munini18}, but it should be noted that the fit is not always applicable \citep{dumbovic11}. This might be related to the interruption by another interplanetary structure or the definition of the recovery phase.

The recovery phase, as defined by most studies, starts at the minimum of the depression, \ie in the two-step FD at the center of the CME magnetic structure. The first part of the recovery phase is therefore governed by the interaction with magnetic ejecta, whereas the exponential decay, related to the shock propagating away, starts after the passage of the magnetic ejecta. Adopting the ``shadow effect'', we assume that the GCR density will recover at an approximately exponential rate, \ie that the GCR density amplitude in this part of the recovery phase can be written as the modified exponential function:

\begin{equation}
A(r)=\frac{U(r)-U_0}{U_0}=A_0(r)\mathrm{e}^{-\frac{r-r_0}{\lambda}},
\label{eq1}
\end{equation}

\noindent where $A(r)$ is the relative amplitude of the GCR count,  $r$ is the radial (heliospheric) distance, $r_0$ is the distance at which the recovery starts, $U(r)$ is the GCR phase space density at distance $r$, $U_0$ is the unperturbed GCR phase space density, $A_0(r)$ is the function which determines the amplitude of the recovery, and $\lambda$ is a constant. Although it has been shown that there is a radial gradient of GCRs of about 3\%/au \citep[\eg][]{webber99,gieseler16,lawrence16}, $U_0$ is assumed to be constant for simplicity reasons. This assumption was used and tested for flux rope-related FDs by \citet{dumbovic18b}, however, the radial gradient of GCRs might have more significant contribution to the recovery phase of a FD, possibly affecting the recovery rate. Note that $A_0(r)$ cannot be constant because $A(r)$ has to satisfy the boundary condition $A(r_{\mathrm{end}})=0$ (\ie that GCR count fully recovers at some distance $r_ {\mathrm{end}}$) and the domain of the exponential function is not restricted. The transformation of Equation \ref{eq1} into the time-series scale can be performed with the substitution $r\rightarrow v_{\mathrm{shock}}t$ (assuming constant shock speed, $v_{\mathrm{shock}}$) yielding the expression:

\begin{equation}
A(t)=A_0(t)\mathrm{e}^{-\frac{t-t_0}{\tau}}\,,\,\,\,\, A_0(t)=A_{\mathrm{sh,max}}+(t-t_0)/\tau,
\label{eq2}
\end{equation}

\noindent where $A(t)$ is the relative amplitude of the GCR count which starts from its maximum value equal to the sheath-related FD amplitude, $A_{\mathrm{sh,max}}$ (see Figure \ref{fig2}), $A_0(t)$ is the function which determines the amplitude of the recovery, $t_0$ the start time of the recovery, and $\tau$ is the characteristic recovery time defined by $\lambda$, $\tau=\lambda/v$. The $A_0(t)$ functional form is chosen arbitrarily, as the simplest form which can satisfy the initial and final state conditions ($A(t_0)=A_{\mathrm{sh,max}}$ and $A(t_{\mathrm{end}})=0$, respectively). We note that the characteristic recovery time, $\tau$, can be treated as a free parameter, although it should be related to the speed and spatial extent of the shock, and include other possible influences, such as the radial gradient of GCRs.

\subsection{The sheath-related effect}
\label{sheath}

The sheath region, on the other hand, has a much larger spatial extent compared to the shock \citep[typically $\sim10$ hours at Earth according to][]{russell02} and is characterised by a highly compressed and fluctuating magnetic field, as well as increased plasma flow. \citet{wibberenz98} describes the sheath-related FD model first proposed by \citet{chih86}, assuming that the sheath acts as a ``propagating diffusive barrier" (PDB). In this 1D model the transport equation \citep{parker65, jokipii71} is reduced to the convection-diffusion equation by adopting the so-called \textit{force field approximation}. The force field approximation assumes a steady state without sources of cosmic rays and neglects the adiabatic energy loss, resulting in a solution given by a parameter describing the rigidity loss called the force field potential \citep{gleeson68,caballero-lopez04}. In the force field approximation the change in the GCR distribution function is given by the exponential of the modulation function, which depends on the flow speed and radial diffusion coefficient \citep[Equations 9 and 11 in ][]{caballero-lopez04}. In the PDB model the sheath is represented by a shell where the flow speed is increased and the diffusion coefficient decreased and both have constant values across the shell (see Figure \ref{fig1}a). The corresponding relative GCR density drop in the sheath (normalised to the onset value) is a linear function of the distance to the border of the shell \citep[for more details see][]{wibberenz98}:

\begin{equation}
 A(r)=\frac{U(r)-U_0}{U_0}=-\frac{V'}{D'}r\,,
\label{eq3}
\end{equation}

\noindent where $U(r)$ is the GCR phase space density at distance $r$ from the border of the shell, $U_0$ is the GCR phase space density at the shell border (the onset value), and $V'$ and $D'$ are the flow speed and the radial diffusion coefficient within the shell, respectively. The maximum amplitude of the relative GCR density drop in the sheath, \ie\, FD magnitude $A_{\mathrm{sh,max}}$ for the shell of thickness $L$ is then given by $A_{\mathrm{sh,max}}=-V'L/D'$. Assuming that the diffusion coefficient relates to the magnetic field strength $D'\approx1/B'$ \citep[see \eg][and references therein] {potgieter13}, the sheath-related FD magnitude will depend on the flow speed in the sheath, the magnetic field strength in the sheath and the sheath thickness. The FD drop rate $\mathrm{d}A(r)/\mathrm{d}r$ is given by the slope of the linear function in Equation \ref{eq3} and can be easily shown to be related to the FD magnitude \citep{forstner20}:

\begin{equation}
\frac{\mathrm{d}A(r)}{\mathrm{d}r}=-\frac{A_{\mathrm{sh,max}}}{L}\,.
\label{eq4}
\end{equation}

Since $A(r)$ is a linear function of $r$, which can be written as $r=V't$, one can easily obtain the time-evolving FD drop rate, $\mathrm{d}A(t)/\mathrm{d}t=\mathrm{d}A(r)/\mathrm{d}r\cdot V'$ \citep{forstner20}. Note that this is the steady-state solution, therefore, the sheath evolution might also lead to changes in the FD profile and/or magnitude. We note that \citet{wibberenz98} used PDB to explain the recovery phase of FDs as well, noting that unlike the main phase, the recovery is determined by the global propagation conditions, \ie\, by the changes at the shock front. However, such an approach implicitly assumes that the disturbed conditions of the transport parameters persist well after the CME passage, whereas we would expect the interplanetary space to return to its undisturbed state. Another strong argument against such approach is the fact that the recovery is well-represented by the exponential term which is not necessarily energy-dependent \citep{lockwood86,jamsen07, usoskin08}. This indicates that the recovery phase primarily depends on the decay of the disturbance and only secondarily on the transport parameters, favouring the so-called ``shadow effect" (although the energy-dependence could be introduced by allowing the shield-effect to be different for particles of different energies). Nevertheless, assuming that the ``shadow effect" is indeed energy-independent, the energy-dependence found in some studies \citep[\eg][]{jamsen07,usoskin08,zhao16b,munini18} might be related to the fact that the exponential recovery phase in those studies was defined to start at the minimum of the depression, \ie in the two-step FD at the center of the CME magnetic structure. The first part of the recovery phase is therefore governed by the particle interaction with the magnetic ejecta, which is energy-dependent. \citet{usoskin08} found that all largest ($>10\%$) FDs demonstrate an energy dependence of the recovery time, while smaller events can demonstrate either energy dependence or the lack thereof. Since the largest FDs are most prominently caused by shock-associated ICMEs, where both shock/sheath and CME magnetic structure are encountered \citep{richardson11a}, the sample of energy dependent recovery events might involve an energy dependent part (due to the CME magnetic structure) and an energy independent part (due to the shadow effect of the shock).

\subsection{The CME magnetic structure-related effect}
\label{FR}

Finally, we regard the second step of FDs corresponding to the CME magnetic structure. In \citet{dumbovic18b} an analytical diffusion--expansion Forbush decrease (FD) model \textit{ForbMod} was presented. The model is restricted to explaining the depression caused by the CME magnetic structure, \ie the flux rope (FR), and the interaction between the particles and the FR, which is described \textit{via} diffusion, while taking into account the fact that the FR expands self-similarly (see Figure \ref{fig1}b). Several representative expansion options, related to the effective change of the axial magnetic flux were considered. In particular, in a force free model \citep[\eg][]{lundquist51} for a circular cross-section FR the axial magnetic flux can be written as $\Phi_{\mathrm{ax}}\sim B_ca^2$, where $B_c$ is the magnetic field in the FR center and $a$ is the FR radius. Assuming that $B_c$ and $a$ change with heliospheric distance following a power-law with indices $-n_B$ and $n_a$, respectively, it is trivial to see that in the case when $n_B=2n_a$ the magnetic flux is conserved. A more general expression can be written in the form $\Phi_{\mathrm{ax}}\sim f(t,x)$, where $x=n_B-2n_a$ determines whether the magnetic flux is conserved ($x=0$), increasing ($x<0$) or decreasing ($x>0$). In \citet{dumbovic18b} solutions for 4 specific expansion types ($x=0$, $x=0.5$, $x=-0.5$, and $x=-1$) were provided and analysed. Based on Equations 12 and 15 in \citet{dumbovic18b} it is possible to find a general solution for an arbitrary expansion type $x$, providing $x\neq-1$ (due to integration rules). The GCR phase space density for particles of  a specific rigidity (without energy dependence) can then be written as:

\begin{equation}
 U(\hat{r},t) = U_0\Bigg(1-J_0(\alpha_1\hat{r})\mathrm{e}^{-\alpha_{1}^{2}f(t)}\Bigg)\,,\, \, \,\, \, \,\, \, \,\, \, \, f(t)=\frac{D_0}{a_0^2}\cdot \Big(\frac{v}{R_0}\Big)^x\cdot\frac{t^{x+1}}{x+1}\,,
\label{eq5}
\end{equation}

\noindent where $U_0$ is the GCR phase space density at the FR surface, $J_0$ is a Bessel function (of the first kind) of the order 0, $\alpha_1$ is a first positive root of $J_0$ (tabulated in tables of Bessel functions), $\hat{r}$ is the radial distance from the FR axis to the outer border of the FR, scaled to the FR radius ($\hat{r}=r(t)/a(t)$), $D_0$, $a_0$ and $R_0$ are the initial diffusion coefficient, radius and height, $v$ is the CME speed (assumed to be constant), and $x=n_B-2n_a$ is the expansion type ($n_B$ and $n_a$ are power-law indices for $B_c$ drop and $a$ increase, respectively). Note that it is assumed that $D\sim1/B_c$ and thus also has power-law behaviour with index $n_B$. The decrease is symmetric and constrained within the borders of the flux rope, with the maximum depression in the center of the flux rope, $A_{\mathrm{FR,max}}=-e^{-\alpha_{1}^{2}f(T)}$, where $f(T)$ is given by Equation \ref{eq5} and $T$ is the transit time to the observer. Since $A(r)$ for the flux rope part of the FD is given by the Bessel function, the drop rate is not constant across the flux rope:

\begin{equation}
 \frac{\mathrm{d}A(\hat{r})}{\mathrm{d}\hat{r}}=\alpha_1J_1(\alpha_1\hat{r})e^{-\alpha_{1}^{2}f(t)}\,,
\label{eq6}
\end{equation}

\noindent where we used the Bessel function property $-\mathrm{d}/\mathrm{d}r(J_{\mu}(r)/r^{\mu})=J_{\mu+1}(r)/r^{\mu}$ (for $\mu\ge0$). It can be easily shown that the maximum drop rate depends linearly on the FD magnitude, $(\mathrm{d}A(\hat{r})/\mathrm{d}\hat{r})_{\mathrm{max}}\sim A_{\mathrm{FR,max}}$.

\begin{figure}
\centerline{\includegraphics[width=0.47\textwidth]{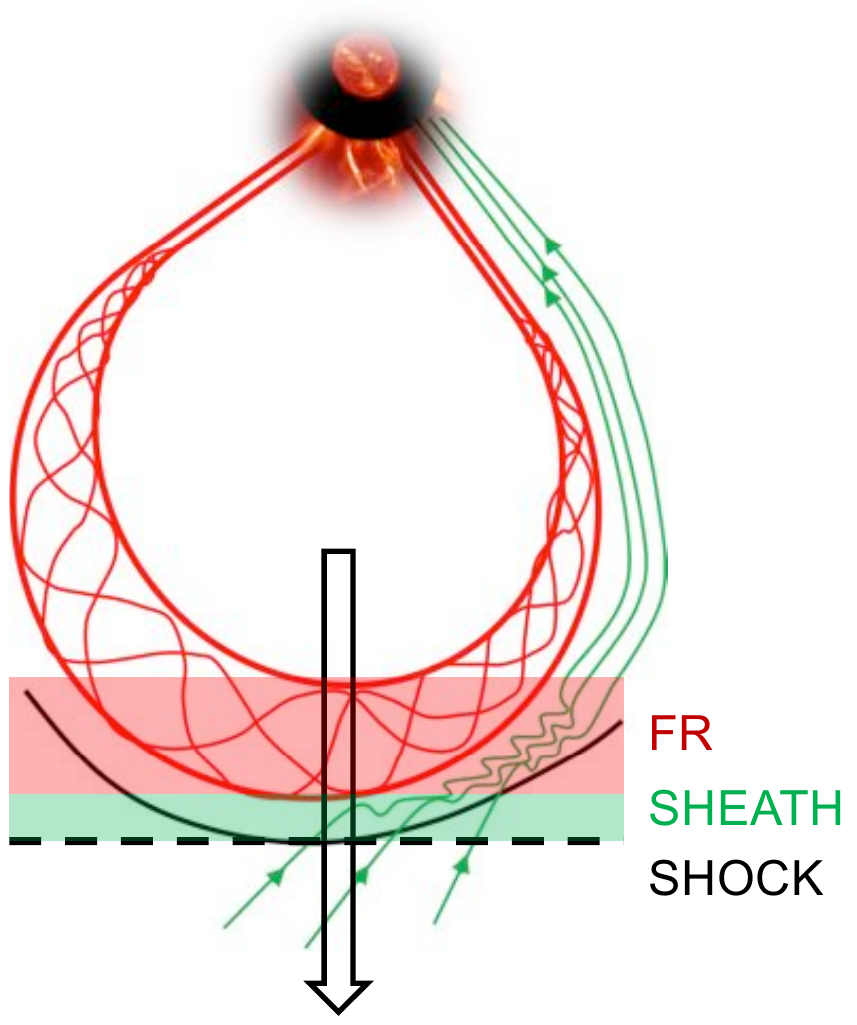}
						\hspace{0.03\textwidth}
		\includegraphics[width=0.47\textwidth]{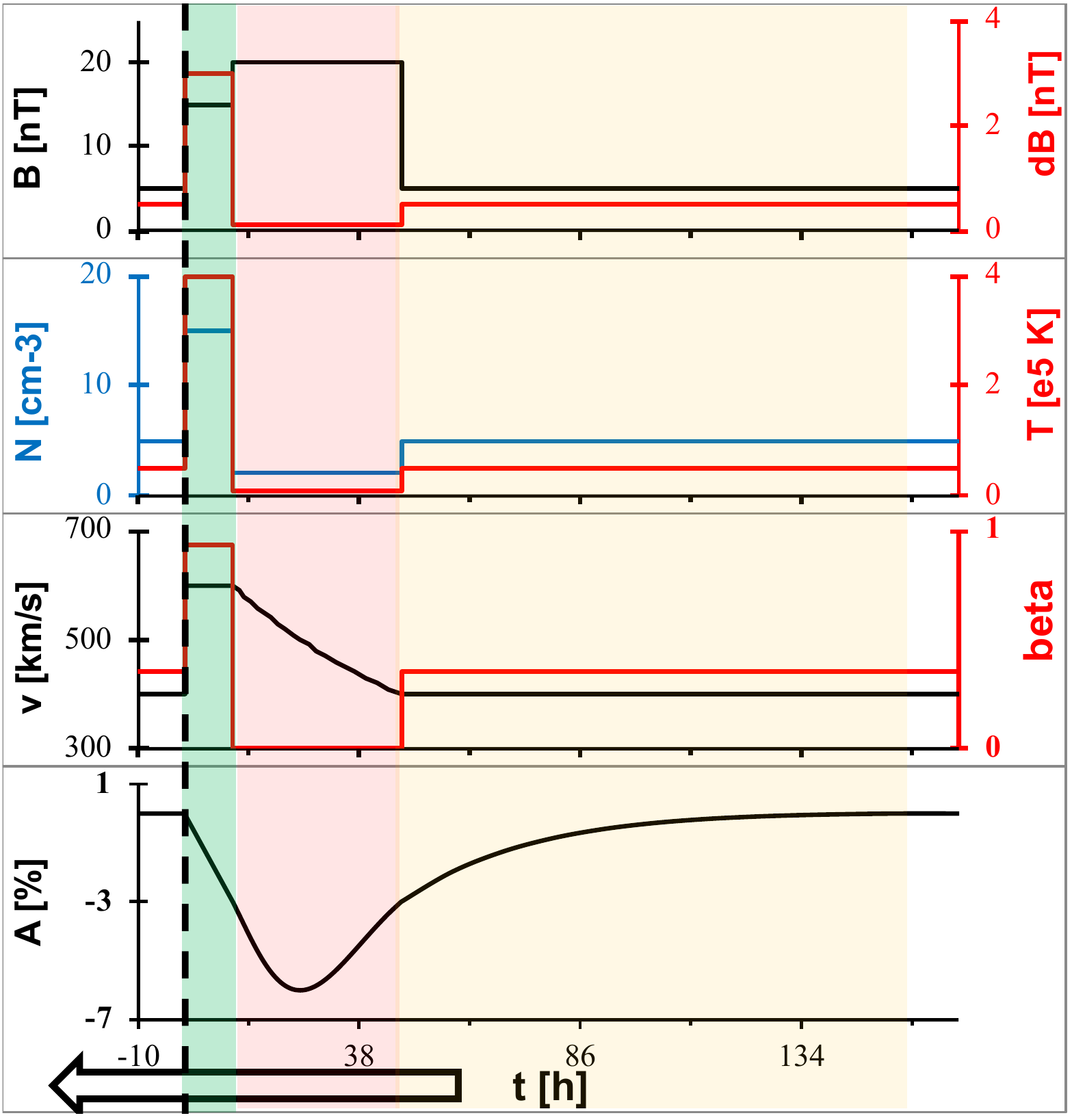}}
\caption{\textit{left:} A sketch of ICME with shock (black line), sheath (green) and CME magnetic structure (red); \textit{right:} sketch of the assumed solar wind plasma parameters, top to bottom: magnetic field strength, $B$ and fluctuations, $\mathrm{dB}$; plasma density, $N$ and temperature, $T$; plasma speed, $v$ and beta parameter; and a generic profile of a two-step FD for particles of energies $>50$ MeV. Shock arrival is marked by the black dashed line, sheath region is highlighted green, CME magnetic structure is highlighted red and the recovery phase is highlighted yellow. The passage of the ICME over the observer is marked by an arrow in both sketches.}
\label{fig2}
\end{figure}

Based on these modelling efforts it can be concluded that different interplanetary structures will interact differently with GCRs resulting in different 'stages' of the FD. These stages can be combined together, by superimposing different effects in  order to obtain a 'generic profile' of a two-step FD. This is shown in Figure \ref{fig2}, along with a sketch of the assumed solar wind plasma parameters. The values of each parameter were selected for each region separately based on typically observed values for magnetic clouds \citep[\eg][]{zurbuchen06,richardson10,richardson11b}. This includes FD amplitudes of the shock/sheath and FR regions, which were normalised to 3\%, following the statistical study of \citet{richardson11b} for particle energies $>60$ MeV. The values of the physical quantities in Figure \ref{fig2} are therefore not necessarily interrelated. Nevertheless, this  aims to show the shape of the FD based on the modelling described above.   Quantitative analysis should involve  real events and will be given in the second part of this paper.

The first stage of the FD is the first step, which starts with the shock arrival and is constrained to the spatial extent of the sheath. The relative amplitude in this region is given by Equation \ref{eq3} and reaches its maximum at the end of the sheath region, with corresponding relative amplitude $A_{\mathrm{sh,max}}$. For simplicity we assume that there is no pre-increase or contributions to the drop due to the shock, \ie that the whole drop in the sheath region is related to the sheath effect and is given by the PDB. The second stage of the FD corresponds to the CME magnetic structure. It  includes not only  the second step of FD, but also the first part of its recovery. It is constrained within the spatial extent of the CME magnetic structure. The relative amplitude is given by Equation \ref{eq5} and reaches its maximum at the center of the CME magnetic structure, \ie  the flux rope, with the corresponding relative amplitude $A_{\mathrm{FR,max}}$. The third and final stage is the exponential recovery, related to the decay of the shock 'shadow'. It starts with the end of the CME magnetic structure,  where the relative amplitude is given by Equation \ref{eq2} and the duration, determined from the condition $A(t_{\mathrm{end}})=0$ is given  by $t_{\mathrm{end}}=t_0-\tau A_{\mathrm{sh,max}}$. The total amplitude of the FD is given by $A_{\mathrm{TOT}}=A_{\mathrm{sh,max}}+A_{\mathrm{FR,max}}$.

Note that the change of the relative amplitude for  the sheath and  the CME magnetic  structures is given across spatial scales, and therefore  needs to be transformed into the time series; that can easily be achieved using the speed profile. Due to the expansion, the speed profile across the FR is linearly decreasing, resulting in an asymmetry in  the time series of the FR-related FD. This asymmetry somewhat 'smears out' the transition points between different regions, even in this highly idealised representation of the FD. Therefore, it is not surprising that the FD was often considered to be a homologous phenomenon, especially given that a number of additional aspects may influence the measured GCR count \citep[\eg time-resolution, energy range, external influences,][]{clem00}, and even the characteristics of the structure \citep[\eg faster ICMEs produce more asymmetric depressions,][]{belov15}. 

\section{Adapting Forbush decrease models for energy-dependence}
\label{energy}

The models used in Section \ref{profile} to reproduce the shape of the two-step FD are not explicitly energy-dependent, however, this is an important feature of FDs. In \citet{dumbovic18b} the case of fixed arbitrary particle energy with a diffusion coefficient which was only a function of time was considered. While this allowed us to qualitatively assess whether the model fits the observations, for a more proper and precise quantitative analysis energy dependence must be considered. This is due to the FD detection limits, as well as additional modulation by \eg planetary magnetic fields, atmospheres or the detectors themselves. The FD magnitude is typically around several percent, therefore, large statistics are needed, \ie high particle count rates. These are easily provided by instruments which count all particles that enter from all directions, regardless of their energy, such as large ground based neutron or muon monitors, as well as single counters onboard spacecrafts.

Detectors which measure particles of a specific energy (or narrow energy intervals) typically provide smaller count rates and thus, in order to observe the effect of several percent, the time resolution needs to be decreased. Moreover, the effect in the low-energy detectors is often masked by the increased flux of low-energy solar energetic particles (SEPs). For example, \citet{munini18} used PAMELA (Payload for Antimatter-Matter Exploration and Light-nuclei Astrophysics) data to analyse the FD recovery in 9 different narrow energy ranges between 0.4 and 20 GV and reported (1) that the statistics allowed time resolution of the proton flux of 3 or 6 hours up to 5 GV and one day above 5 GV; (2)  that the main phase of FD was not visible in $<2$ GV protons due to SEPs. For comparison, during solar minimum SOHO/EPHIN detector F and a typical neutron monitor at the pole have a counting rate of more than 1000 counts/minute \citep{moraal00,kuhl15}, providing sufficient statistics to observe FDs at a minute resolution. We do note that the long integration time needed for PAMELA data is related to its orbit, because the instrument is located at a low-orbiting satellite spending most of the time inside the geomagnetic field and can thus measure low-energy particles only when traversing (sub)polar regions, while SOHO/EPHIN and polar NMs are exposed to low-energy particles at all times. We also note that due to the low cutoff energy (50 MeV), the SOHO/EPHIN detector F also has a problem with the increased flux of SEPs masking the effect, whereas for neutron monitors this will be the case only for very rare, most energetic SEPs which produce ground level enhancements. Since FDs are typically measured with instruments which observe particles of a specific type at all energies above some specific energy/rigidity cutoff, any quantitative comparison of the modelled and observed FDs should consider energy-dependence.

We introduce energy dependence by allowing the diffusion coefficient, $D$, to be a function of rigidity as well as time, which can be expressed through a typical empirical expression as used in numerical models fitted to GCR measurements, as given by \citet{potgieter13} and discussed in detail in Appendix \ref{diff_coeff}. While this expression might be suitable for the diffusion coefficient within a flux rope at Earth, the flux rope will have a different diffusion coefficient at other heliospheric distances, because it is a function of time. It can be extrapolated back and forth in time assuming that the diffusion coefficient time dependence is defined by that of the magnetic field (power-law), \ie $D_0=D(R(t)/R_0)^{-n_B}$, where $D_0$ is the initial diffusion coefficient, $R_0$ and $R(t)$ are different heliospheric distances and $n_B$ is the expansion factor (power-law index) of the magnetic field strength.

It can be easily shown that allowing the diffusion coefficient to be energy-dependent does not affect the radial dependence in either PDB or \textit{ForbMod} (because the diffusion coefficient appears only in the time-dependent part), therefore, the initial assumptions are not violated with the introduction of an energy-dependent diffusion coefficient. Addition of the energy-loss effect (adiabatic cooling) on the other hand would change the starting assumptions in both models and is thus not formally taken into account within the models. We note that on the heliospheric scale adiabatic cooling is relevant mostly for lower energy particles \citep[$E<100$ MeV;][]{gleeson71}, although it might have larger impact for highly expanding CMEs. In general, the energy-loss is expected to introduce additional modulation effects \citep[see \eg][and references therein]{lockwood71}, but the exact quantitative contribution is not trivial to estimate. From the qualitative aspect it is expected that the energy--loss term acts to balance out the inward diffusion of particles \citep{munakata06} and therefore acts to increase FD amplitude. 

Note that the solution given in Equation \ref{eq5} is only valid for a specific rigidity $P$, where a more general GCR phase space density would be given as $U(\hat{r},t, P)$, with $U_0 \rightarrow U_0(P)$, $D_0 \rightarrow D_0(P)$ and consequently $A(r,t) \rightarrow A(r,t,P)$. The total GCR phase space density in the FR after time $t$ and at distance $\hat{r}$ is obtained by integrating over all available rigidities. Since we are interested only in detected particles, the phase space is constrained by the detector cutoff, $P>P_{\mathrm{cut}}$. Recognising that $\int_{_{P>P_{\mathrm{CUT}}}}^{} U_0(P)dP$ represents the total GCR phase space density at the FR surface for all rigidities that can be measured by the detector, \ie the total quiet-time GCR phase space density $U_{0,TOT}$, the total FD amplitude can be written as:

\begin{equation}
A(\hat{r},t)=\int \frac{U(r,t,P)-U_0(P)}{U_0(P)}\mathrm{d}P=-\frac{\int_{_{P>P_{\mathrm{CUT}}}}^{} U_0(P)J_0(\alpha_1\hat{r})\mathrm{e}^{-\alpha_{1}^{2}f(P,t)} \mathrm{d}P}{U_{0,TOT}}\,,
\label{eq7}
\end{equation}

\noindent where $f(P,t)$ is given by the same expression as in Equation \ref{eq5}, except that $D_0$ is a function of rigidity, $D_0(P)$. On the other hand, from Equation \ref{eq5} it can easily be derived that the FD amplitude for a particle of specific rigidity can be written as $A(P)=-J_0(\alpha_1\hat{r})\mathrm{e}^{-\alpha_{1}^{2}f(P,t)}$. Therefore, the total FD amplitude can be expressed as:

\begin{equation}
A(\hat{r},t)=\int_{P>P_{\mathrm{CUT}}}^{} U_{\mathrm{FRAC}}(P)\cdot A(P)\mathrm{d}P\,,\, \, \,\, \, \,\, \, \,\, \, \, U_{\mathrm{FRAC}}(P)=\frac{U_0(P)}{U_{0, TOT}}\,,
\label{eq8}
\end{equation}

\noindent where $U_{\mathrm{FRAC}}(P)$ can be interpreted as the contribution of different energy particles to the total FD. It can be easily shown that Equation \ref{eq8} is valid for the sheath-related FD as well, where FD amplitude for a particle of specific rigidity can be written as $A(P)=-V'r/D'(P)$.

The expression given in Equation \ref{eq8} refers to the amplitude calculated based on the particle phase space density, whereas our aim is to compare it with the amplitude measured by the detector which relates to the GCR count rate in the direct space. For a specific detector, the GCR count rate will depend on the particle intensity, \ie the GCR spectrum and the yield function of the detector, $C=\int J(E)\cdot Y(E)\mathrm{d}E$ \citep[see \eg][]{sullivan71,clem00}, where $J(E)$ is the GCR spectral intensity and $Y(E)$ is the function describing the detection response of the detector, as well as any other influence (\eg from the planetary atmosphere). Note that due to simplicity, in the analytical models presented in Section \ref{profile} and thus also here, we regard only one particle species, \ie protons. Since energy-dependent intensity spectra are the same as rigidity-dependent density spectra \citep[up to a normalisation factor, see][]{moraal13} Equation \ref{eq8} can be rewritten as:

\begin{equation}
A(\hat{r},t)=\int_{E>E_{\mathrm{CUT}}}^{}\xi(E)\cdot A(E)\mathrm{d}E\,,\, \, \,\, \, \,\, \, \,\, \, \, \xi(E)=\frac{J(E)\cdot Y(E)}{\int_{E>E_{\mathrm{CUT}}}^{} J(E)\cdot Y(E)\mathrm{d}E}\,
\label{eq9}
\end{equation}

\noindent where $\xi(E)$ represents the fractional contribution of different energy particles to the total FD.

\begin{figure}
 \centerline{\includegraphics[width=0.99\textwidth]{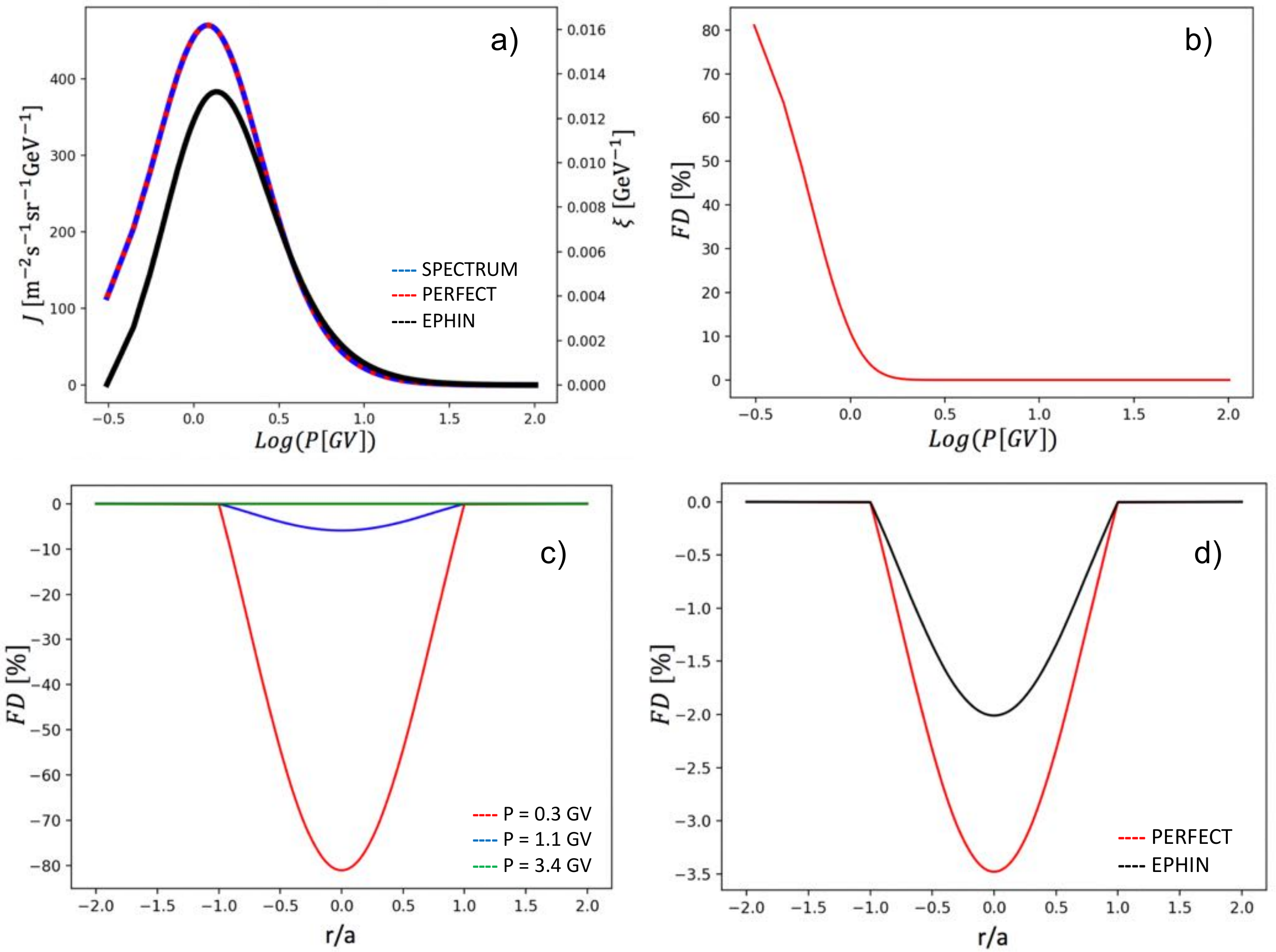}}
\caption{a) GCR spectral intensity obtained using the force--field approximation (SPECTRUM) and the GCR fractional contribution, $\xi$, calculated based on Equation \ref{eq9} for a "perfect detector`` (PERFECT) and SOHO/EPHIN (EPHIN); b) \textit{ForbMod} FD amplitude rigidity dependence; c) \textit{ForbMod} FD profiles for several selected rigidities; d) \textit{ForbMod} total FD integrated over all possible energies for a "perfect detector`` (PERFECT) and SOHO/EPHIN (EPHIN). The exemplary event is based on \insitu measurements presented in Figure \ref{fig2} with an assumed magnetic field expansion factor of $n_B=1.8$ (for details see main text).}
\label{fig3}
\end{figure}

The GCR spectral intensity can be obtained using the so--called ``force--field" approximation \citep[\eg][and references therein]{gleeson68,caballero-lopez04,herbst10,usoskin11,gieseler17}, as discussed in detail in Appendix \ref{force_field}. Figure \ref{fig3}a shows the GCR spectral intensity obtained using the force--field approximation (see Appendix \ref{force_field} for details on the calculation) for February 2014 (the timeframe is chosen to be in line with the real event analysed in Section \ref{obs}). In the same figure we overlay plots of the calculated GCR fractional contribution, $\xi$, for a ``perfect detector" and a real detector (the single detector F of the SOHO/EPHIN instrument). A ``perfect detector" responds to all energies above 0.05GeV equally (\ie $Y(E)=1$), and therefore simply gives a scaled spectral intensity, whereas SOHO/EPHIN responds better to high-energy particles (for details see Appendix \ref{ephin}). We can see that not all energies contribute equally to the observed GCR count rate and that the main contribution is coming from a quite narrow energy range, as was discussed earlier by \citet{rodari18} and \citet{dumbovic19} (note that $J$ and $\xi$ are not given in logarithmic scale). This is the case even for the ``perfect detector" and even more so in the case of the SOHO/EPHIN, due to its energy-dependent response.

In Figures \ref{fig3}b and c we show the rigidity dependence of the FD amplitude and FD profiles for three selected energies, respectively, calculated using \textit{ForbMod}. As a demonstration of the theory we use the event shown in Figure \ref{fig2}, representing a typical magnetic cloud example with the following values needed as input for \textit{ForbMod}: propagation speed $v=500$ \kmps, FR radius $a=0.15$ au, central magnetic field strength at Earth, $B=20$ nT. The size expansion factor, $n_a=0.91$, was calculated from the linear speed profile similar to \citet{demoulin09}, yielding an initial FR radius $a_0=2.9$ \Rsun\,for the initial heliospheric distance of $R_0=15$ \Rsun\,(with corresponding transit time $TT=77$ h). The magnetic field expansion factor $n_B=1.8$ was chosen as a typically expected value \citep[\eg][]{leitner07}. We see a quite strong rigidity dependence of the FD amplitude, although it should be noted that this rigidity dependence, is calculated for monoenergetic cases and is therefore not directly comparable to the observed FD rigidity dependence which is measured for different energy ranges \citep[for details on the observation of FD rigidity dependence see overviews by][]{lockwood71,cane00}. Finally, in Figure \ref{fig3}d we show the total FD for the exemplary event, integrated over all possible energies for the  ``perfect detector" and SOHO/EPHIN. We can see that the FD magnitude for SOHO/EPHIN is almost two times smaller than for the ``perfect detector", the difference coming from the fact that SOHO/EPHIN, unlike the ``perfect detector", has an energy-dependent response. The FD magnitude calculated for the ``perfect detector" can be therefore taken as a very rough upper limit (\ie we always expect to observe smaller FD magnitude in a real detector compared to the ``perfect detector"). The FD magnitude calculated for the ``perfect detector" ($\sim3.5\%$) is somewhat smaller, but still comparable to an average total FD amplitude observed by the IMP 8 spacecraft near Earth \citep[4.3\%,][]{richardson11b}. It should be noted that the total FD amplitude reported by \citet{richardson11b} includes both the sheath and FR contribution, indicating that the value calculated for SOHO/EPHIN ($\sim2\%$) is relatively close to the average FR-related FD amplitude observed by the IMP 8 spacecraft near Earth. However, it should be taken into account that the exemplary event is a strong magnetic cloud, where FDs larger than average are usually observed \citep{richardson11b}.

\section{Multi-spacecraft observation of the February 2014 event}
\label{obs}	

\subsection{In situ measurements at Earth}
\label{insitu}	

In order to fully understand and test the analytical models related to CME-GCR interaction, multi-spacecraft measurements are needed, obtained from radially aligned spacecraft at different heliospheric distances (corresponding to different evolutionary stages of a CME). For that purpose we utilise a study by \citet{winslow18} who studied a single CME, launched from the Sun on 2014 February 12, and its corresponding \insitu signatures at Mercury, Venus, Earth and Mars, including GCR measurements at Mercury, Earth and Mars. In order to analyse this event at Earth, in addition to the measurements used by \citet{winslow18}, we include spacecraft ion composition and suprathermal electron measurements, as well as spacecraft GCR measurements. These can help us to determine more reliably different sub-structures within one ICME event. For that purpose we use data from the following spacecraft/instruments:
\begin{itemize}
 \item \textit{Magnetic Field Instrument} \citep[MFI,][]{lepping95} onboard \textit{Wind}: GSE components of the magnetic field, $B_i$, magnetic field strength, $B$ and fluctuations, $dB$
 \item \textit{Solar Wind Experiment} \citep[SWE,][]{ogilvie95} onboard \textit{Wind}: GSE components of the flow speed, $v_i$, and total flow speed, $v$, plasma density, $N$ and temperature $T$ (with expected temperature calculated according to \citet{lopez87} and \citet{richardson95}), azimuthal flow angle and plasma beta, and ion composition
 \item \textit{Solar Wind Ion Composition Spectrometer} \citep[SWICS,][]{gloeckler98} onboard the \textit{Advanced Composition Explorer} (ACE): plasma composition
 \item South Pole Neutron Monitor GCR count measurements (relative counts) obtained from the Neutron Monitor Database (NMDB) search tool \url{http://www.nmdb.eu/nest/}
 \item GCR count measurements obtained from the \textit{Electron Proton Helium Instrument} \citep[EPHIN,][]{muller-mellin95} onboard the \textit{Solar and Heliospheric Observatory} (SOHO)
\end{itemize}

\begin{figure}
\centerline{\includegraphics[width=0.75\textwidth]{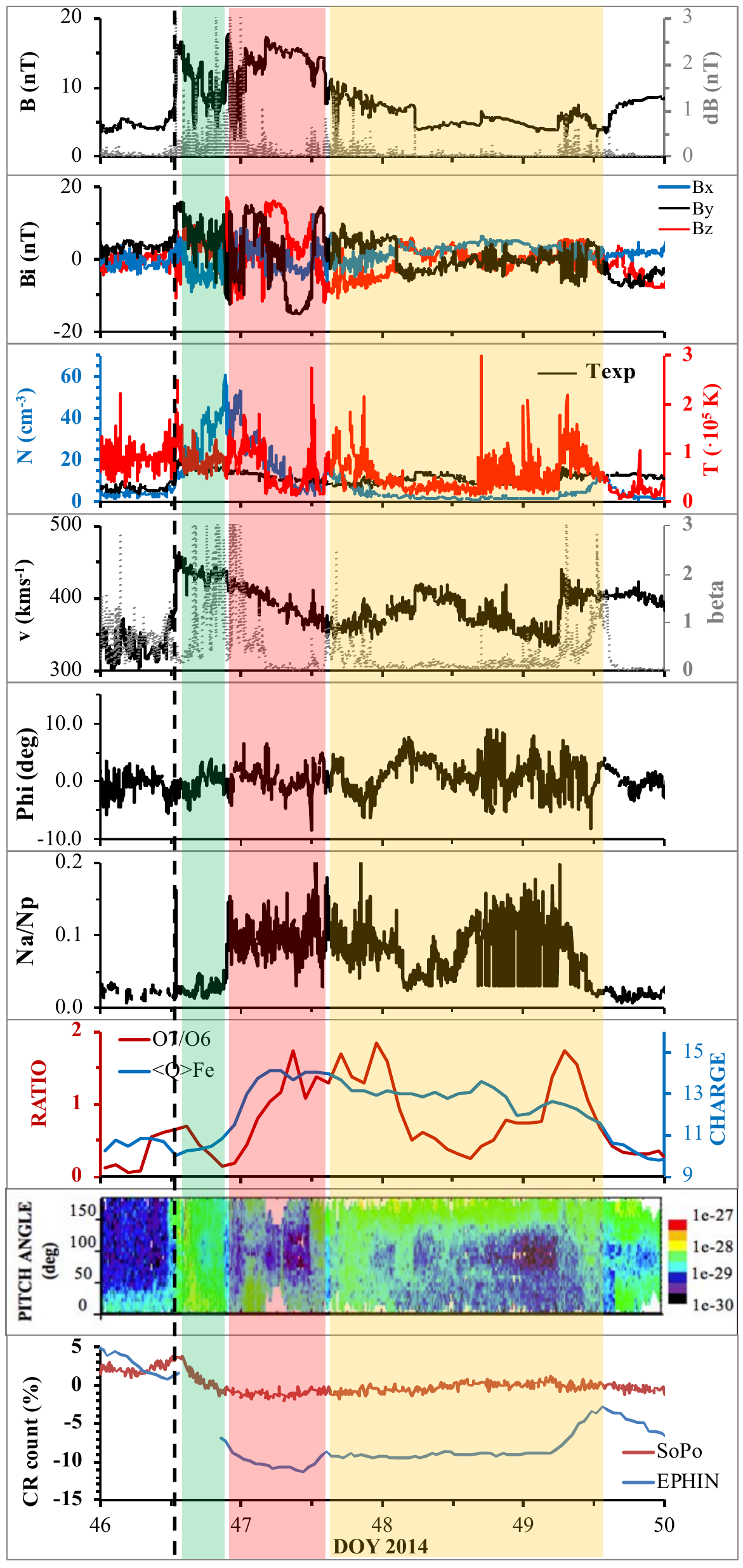}}
\caption{\textit{In situ} measurements for the 2014 February 15 ICME observed near Earth. Different panels show (top to bottom): (1) magnetic field strength, $B$, and fluctuations, $dB$; (2) GSE components of the magnetic field, $B_i$; (3) plasma density, $N$, and temperature, $T$;  (4) plasma flow speed, $v$, and beta; (5) plasma flow angle; (6) alpha to proton ratio, Na/Np; (7) Iron charge states, $<\mathrm{Q}>$Fe, and oxygen charge states ratio, O7/O6; (8) suprathermal electron pitch angle distribution (for E=116,1 eV); (9) relative CR count for South Pole neutron monitor (geomagnetic rigidity cutoff 0.10 GV, altitude 2820 m) and SOHO/EPHIN F-detector ($>$50 MeV). The black line marks the shock and the sheath is shaded green. The region containing magnetic cloud signatures is shaded red and the region with compound stream signatures is shaded yellow, where the borders of both regions are defined based on the 2nd and 3rd depression as observed by SOHO/EPHIN.
}
\label{fig4}
\end{figure}

In Figure \ref{fig4} we present \insitu measurements for the February 15 ICME observed at Earth, which is the interplanetary counterpart of the February 12 CME. We note that inclusion of additional \insitu measurements compared to that used by \citet{winslow18}, especially of the SOHO/EPHIN instrument, somewhat changes the perspective on the event. We highlight three distinct regions we observe in SOHO/EPHIN. The first is the region where the F-detector of the SOHO/EPHIN instrument is dominated by low-energy particles (we remove this data so that the measurement scale is suitable for the small depression to be observed). The second region is characterised by a small, relatively symmetric depression constrained in a time period of linearly declining flow speed profile, increased levels of Fe charge states, as well as O7/O6 and alpha-to-proton ratios, and increased magnetic field. The back of region 2, shows ordered and smooth magnetic field properties and counterstreaming electrons indicating a well defined, twisted and closed magnetic structure. In the front part of region 2 we observe increased magnetic fluctuations, plasma beta, temperature and density, all of which indicate a sheath region, however, the composition and flow speed seem to be connected with the smooth and ordered structure in the back of the region 2 indicating that they belong to the same structure. These frontal mixed plasma/ICME signatures can be interpreted as a consequence of flux rope erosion at its frontal part by the interaction with the solar wind. However, from the GCR point of view, it would seem that although the interaction disturbed one part of the structure, GCRs still perceive it as a single global structure. Region 3, highlighted yellow in Figure \ref{fig4}, presents a new, compact structure as seen from the GCR behaviour in SOHO/EPHIN. However, we note that other \insitu measurements show complex signatures. Low plasma beta, smooth magnetic field and composition indicate a CME-like magnetic structure, but other parameters show a complex structure typical for complex ejectas/compound streams \citep{burlaga03,lugaz17}.

\citet{winslow18} identified somewhat different borders for the ICME, but we note that they used cosmic ray measurements from Cosmic Ray Telescope for Environmental Radiation (CRaTER) on the Lunar Reconnaissance Orbiter \citep[LRO;][]{spence10} and the South Pole neutron monitor (SoPo), where only one drop of prolonged duration is observed, without fine substructures as observed by SOHO /EPHIN. This is probably due to the fact that SOHO/EPHIN is sensitive to lower energy particles than CRaTER and SoPo. Moreover, since they did not observe a second decrease in the GCR data, \citet{winslow18} concluded that the ejecta is probably already filled with GCRs by the time it reaches Earth. Our observations are in agreement with their conclusion, as we only observe a second decrease in a detector responsive to lower energies (SOHO/EPHIN), whereas no additional decrease is observed in SoPo, which is probably directly related to the energy dependence shown in Section \ref{energy}.

\subsection{Remote sensing}
\label{remote}	

\begin{sidewaysfigure}
\centerline{\includegraphics[width=1\textwidth]{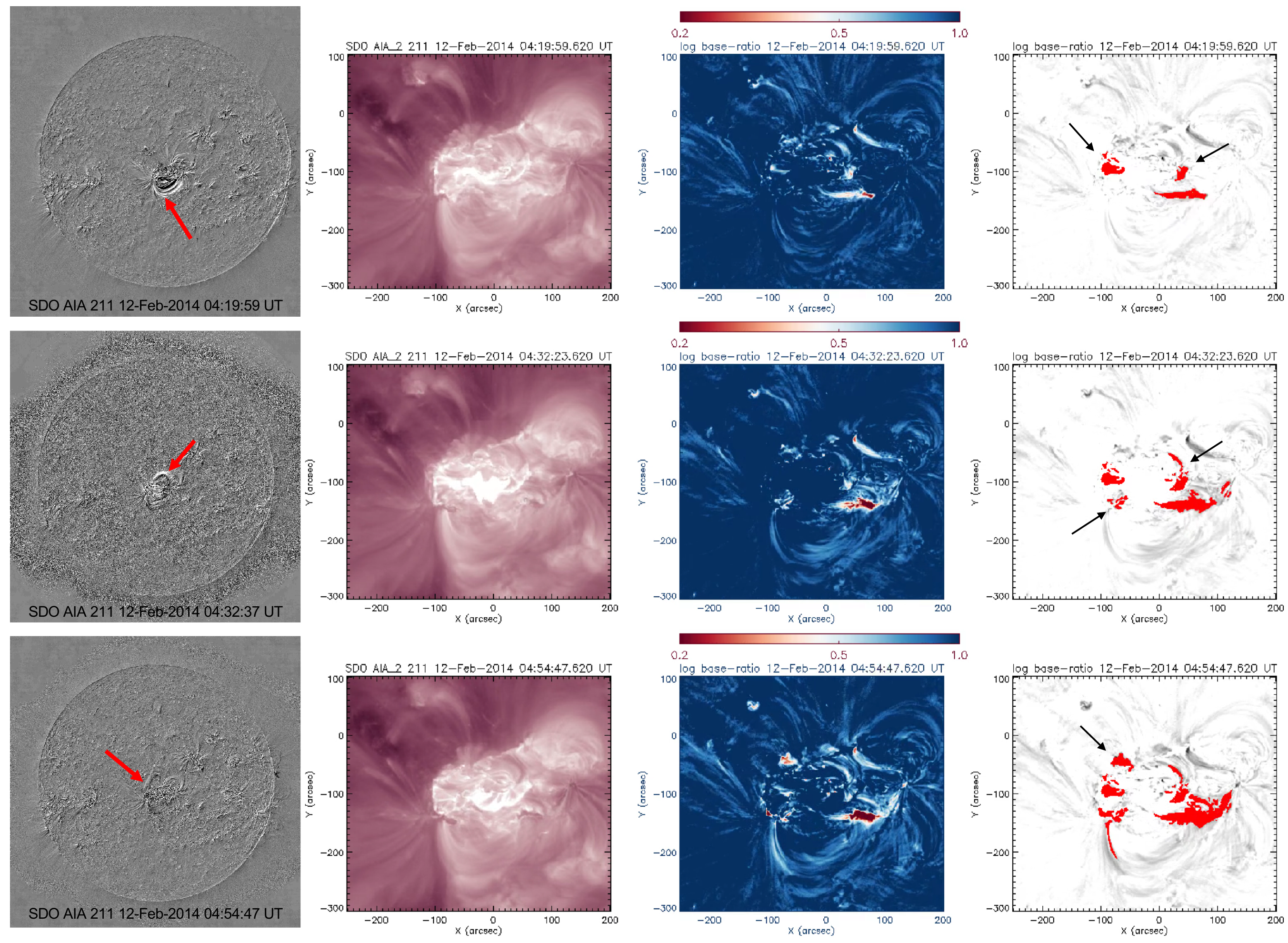}}
\caption{Low coronal signatures of the 2014 February 12 CME: a multi-step eruption is observed with three sub-eruptions from the same active region followed by dimmings at $\sim$04:20 UT (upper panels), $\sim$04:30 UT (middle panels), and $\sim$04:55 UT (bottom panels). On the left: full disc images of SDO AIA 211 in running difference are presented with arrows pointing at the eruptive loop. On the right: a zoom-in on the active region is shown in the same wavelength, together with log-base ratio and highlighted dimmings in the log-base ratio, where the black arrows point to dimmings corresponding to different stages of the eruption (for detailed explanation see main text).}
\label{fig5}
\end{sidewaysfigure}

\begin{figure}
\centerline{\includegraphics[width=0.95\textwidth]{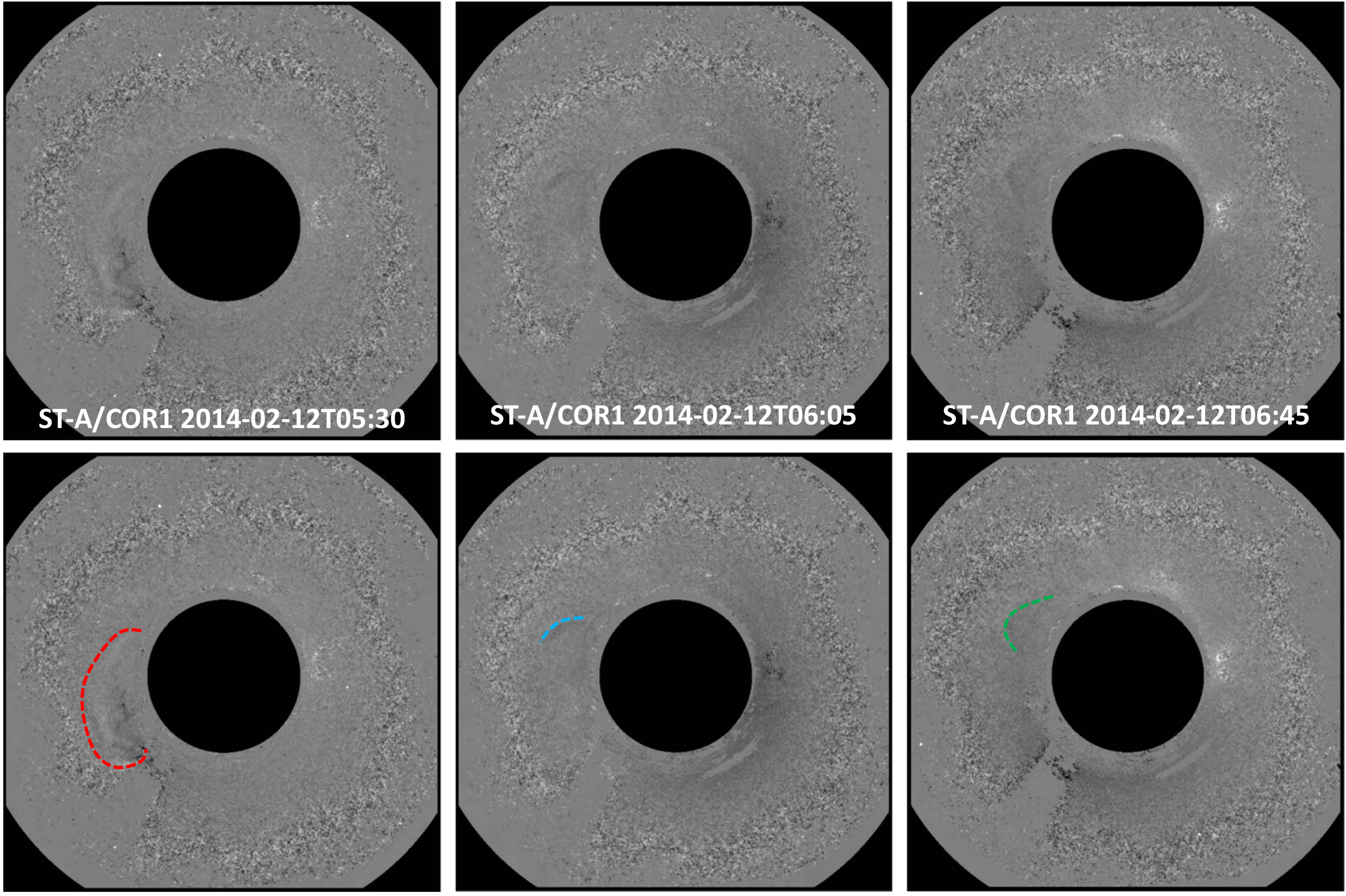}}
\vspace{0.03\textwidth}
\centerline{\includegraphics[width=0.95\textwidth]{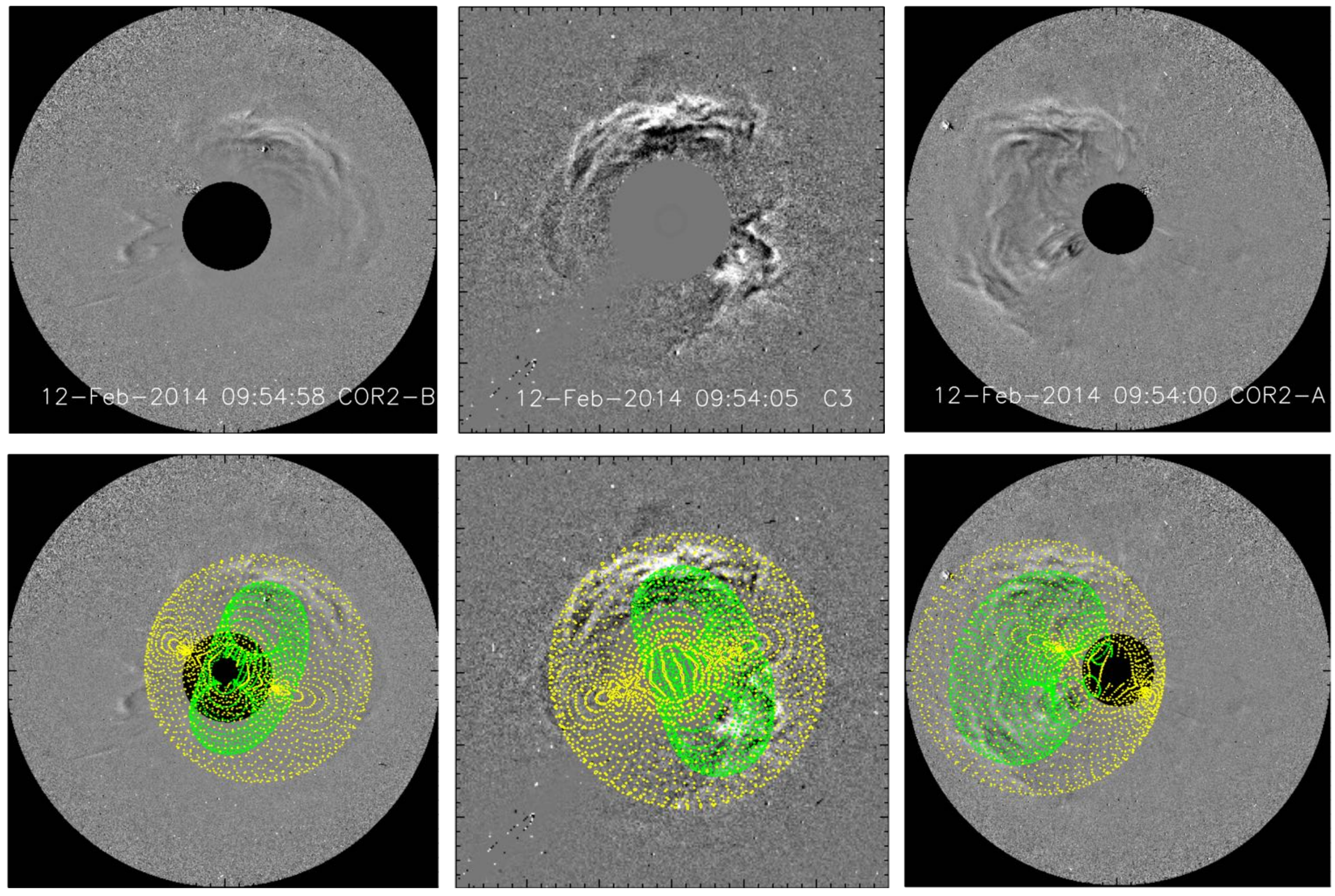}}
\vspace{0.03\textwidth}
\centerline{\includegraphics[width=0.99\textwidth]{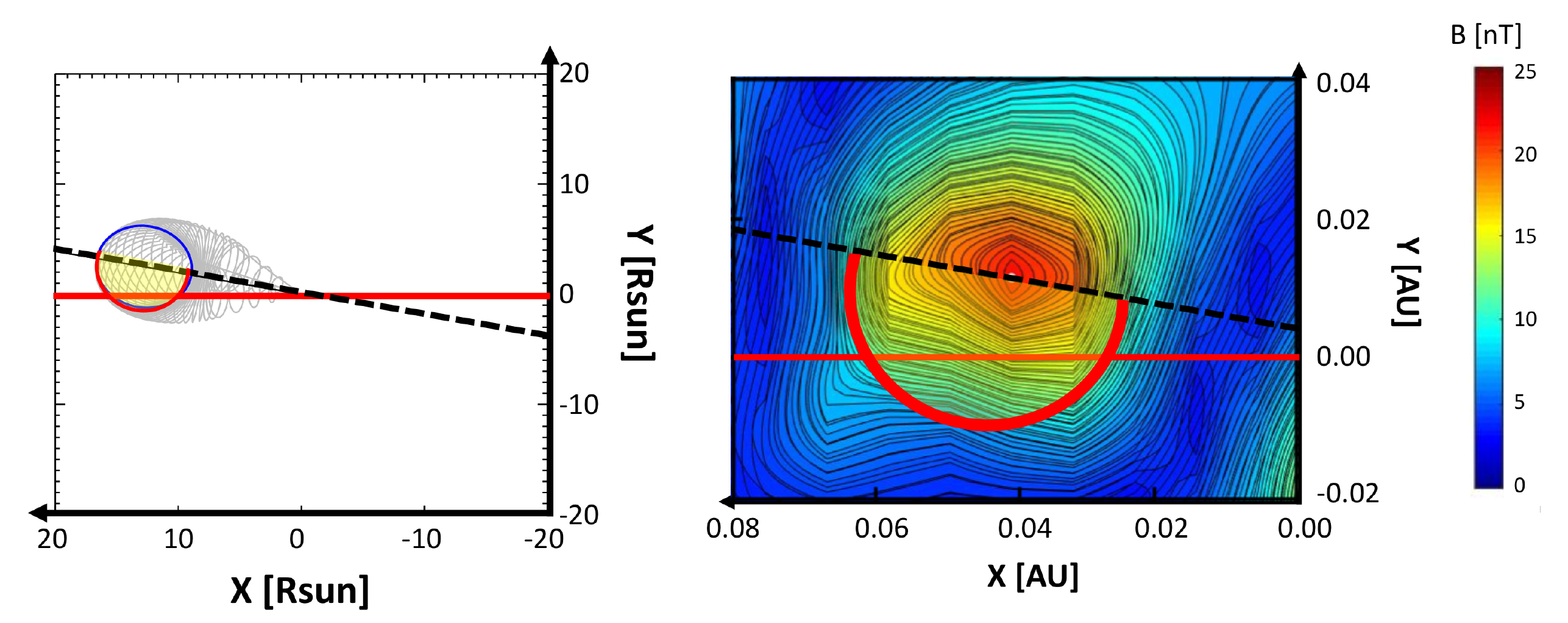}}
\caption{Three eruptions observed at three different times in ST-A COR1 (upper panels); GCS reconstruction of the CME corresponding mainly to the first eruption and of the shock, using STEREO A\&B/COR2 and LASCO C3 (middle panels); comparison of the FR orientations obtained by GCS reconstruction and Grad-Shafranov reconstruction (left and right in the lower panels, respectively). For further details see main text.}
\label{fig6}
\end{figure}

In order to better understand the observed \insitu measurements, we perform a more detailed study of its solar sources. To this aim we investigate coronagraph images of the CME using the SOHO/LASCO  \citep{brueckner95} coronagraphs C2 and C3, and STEREO/SECCHI \citep{howard08} coronagraphs COR1 and COR2, as well as its on-disk low coronal signatures using the \textit{Atmospheric Imaging Assembly} \citep[AIA,][]{lemen12} EUV imager onboard the \textit{Solar Dynamics Observatory} (SDO). We analyse the eruption as seen in SDO AIA 211 $\mathrm{\AA}$ and the corresponding coronal dimmings which are detected based on a thresholding technique applied to logarithmic base-ratio images \citep{dissauer18a}. The eruption site is active region AR11974, a quite large and complex AR, where several smaller eruptions can be observed just before a CME is detected in the STEREO-A COR1 field of view. The first eruption occurs around 04:20 UT, where the eruptive loops are observed moving in south-west direction away from AR11974 (see upper panels of Figure \ref{fig5}). Two core dimmings can be associated with this eruption; these presumably mark the footpoints of the twisted magnetic structure \citep{hudson96,mandrini05} with the axis approximately aligned with the solar equator. Thereafter, a secondary dimming appears south-west of the eruption site, in the direction in which the eruptive loops were observed to propagate. At $\sim$ 04:30 UT a second eruption is observed with the eruptive loops moving away from the eruption site in north-west direction, followed by the appearance of two new dimmings whose connecting line is tilted by $45^{\circ}$ with respect to the solar equator. Finally, at around 04:55 UT a third eruption is observed with the eruptive loops moving away from the AR in the north-east direction and a pronounced dimming is detected north of the AR, whereas the dimming area to the south-west is growing. The three eruptions are followed by a CME detected in STEREO-A COR1 at around 05:05, moving in the south-east direction and expanding as a bean-shaped front (left upper panel in Figure \ref{fig6}). There are two additional faint fronts appearing at 05:40 UT and 06:35 UT around the same position angle, but directed slightly more to the north (middle and right panel in Figure \ref{fig6}). These three fronts are very likely to correspond to the three eruptions observed by SDO. In COR2 and LASCO the three components are not distinguishable, especially in LASCO where a single halo CME is observed. The three eruptions are most likely interrelated and possibly form a single CME. However, it is reasonable to assume that such a CME would not show a nice ordered flux rope structure in \insitu measurements and would very likely result in compound stream signatures.

We perform a 3D CME reconstruction using the Graduated Cylindrical Shell model \citep[GCS,][]{thernisien06,thernisien09,thernisien11} which assumes that geometrically the magnetic structure of a CME can be represented as a hollow croissant with its origin at Sun center, \ie with conical legs, circular cross section and pseudo-circular front. At a specific time the croissant is fully defined by the position of its apex (latitude, stonyhurst/carrington longitude, height), tilt (orientation of its axis with respect to solar equator), half-angle (the angle between the central axis of the legs) and ratio (parameter defining the thickness of the conical legs). We fit the projection of the croissant to coronagraphic images from the STEREO-A and -B/COR2 and LASCO/C3 coronagraphs (\ie from three different vantage points) recorded at approximately the same times to better constrain the fit. The fits are done manually with the main constraint being the structure observed in ST-A as a continuation of the first eruption detected in ST-A/COR1 (left upper panel in Figure \ref{fig6}). We obtain the best fit for the following GCS parameters: $11^{\circ}$ longitude, $-6^{\circ}$ latitude, $-70^{\circ}$ tilt, 0.3, ratio and $28^{\circ}$ halfangle. The GCS reconstruction of the CME at a height of 17\Rsun is given by the green mesh in the middle plots of Figure \ref{fig6}, whereas the yellow mesh represents the reconstruction of the corresponding shock, obtained assuming it has a spherical shape and similar tilt/source position as the CME. We perform GCS at several different time-steps to estimate the speed of the CME apex ($v=800$\kmps).

We compare the orientation obtained from the GCS reconstruction with the orientation of the magnetic structure corresponding to the 2nd step of the FD observed by SOHO/EPHIN, as obtained by Grad-Shafranov (GS) reconstruction \citep{hu02,mostl09,hu17}. GS reconstruction is a 2.5D numerical method to obtain flux rope orientation and magnetic field based on solving of the GS equation in the De Hoffmann-Teller frame (frame of static flux rope) and optimal fitting to the measurement data. As a result one derives orientation and radius of the flux rope, as well as the magnetic field contour plot in the xy-plane of the spacecraft, where the goodness of the fit is determined by the minimum of fit residuals, $R_f$ \citep[with $R_f<0.2$ being the condition for a satisfactory solution, see][]{hu17}. We note that although the best reconstruction ($R_f=0.11$) is obtained only for the very inner part of the FR where a clear rotation is observed (we do not present these results here), for the borders defined based on the 2nd step of the FD observed by SOHO/EPHIN we find a borderline ($R_f=0.2$) solution.

The comparison of GCS and GS reconstruction is shown in the bottom panels of Figure \ref{fig6}. The left bottom panel of Figure \ref{fig6} shows the GCS reconstruction projected onto the solar equatorial plane, with the black dashed line showing the direction of the apex, the red line showing the Sun-Earth line and the red semicircle outlining the cross-section of the croissant in the solar equatorial plane. The GCS results are overlaid on the GS reconstruction image, as determined by the GS method, showing a magnetic field contour plot in the plane of the flux rope cross section which is practically perpendicular to the solar equatorial plane (in the right bottom panel of Figure \ref{fig6}). The agreement between the orientation of the two reconstructions, can be seen from the tilt angle agreement (GCS tilt angle is $-70^{\circ}$ and GS tilt angle is $-75^{\circ}$). This agreement indicates that the GCS reconstruction fits the main part of the CME, which shows ordered magnetic structure in the \insitu measurements corresponding to the 2nd step of FD observed by SOHO/EPHIN. We note that both the GCS and GS reconstruction suggest that the Sun-Earth direction is not perfectly aligned along the diameter of the flux rope, however, the diameter obtained by GS reconstruction ($0.15$ au) is in good agreement with the measured FR size ($0.16$ au, see Table \ref{tab1}).

\subsection{Multi-spacecraft in situ measurements}
\label{multi-spacecraft}	

\begin{figure}
\centerline{\includegraphics[width=0.62\textwidth]{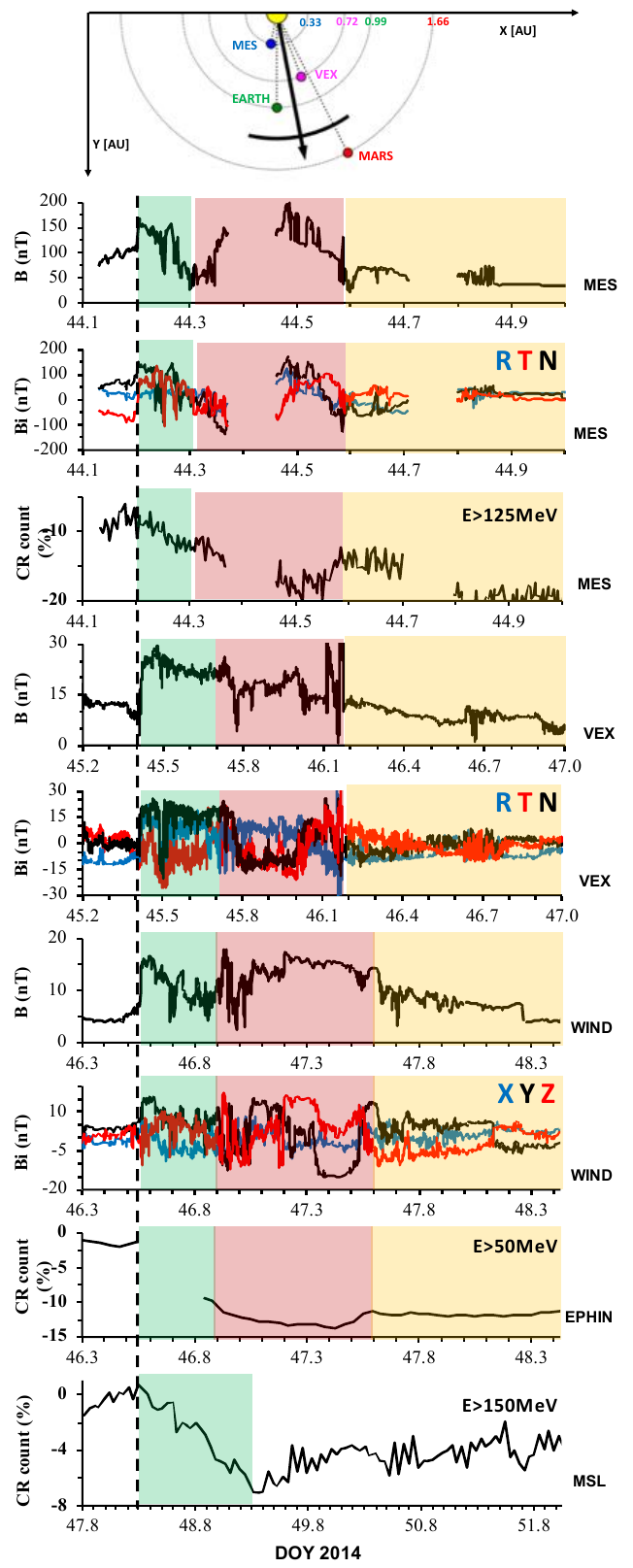}}
\caption{Spacecraft positions (top; CME direction and angular extent are marked in black) and multi-spacecraft \textit{in situ} measurements for the 2014 February 12 CME (from top to bottom): magnetic field strength, RTN components of the magnetic field and GCR counts at \textit{Messenger} (MES, first three panels); magnetic field strength and RTN components at \textit{VEX} (panels 4 and 5); magnetic field strength and GSE components at \textit{Wind} and GCR counts at \textit{SOHO/EPHIN} (panels 6-8); and GCR counts at \textit{MSL/RAD} (bottom panel). The dotted line marks the shock arrival time at each spacecraft and the sheath is highlighted green. The two magnetic structures as identified at Earth from Figure \ref{fig4} are identified at \textit{MES} and \textit{VEX} and highlighted red and yellow, respectively. (Note that the time scales at different spacecraft are not the same and the second magnetic structure is not encompassed at all spacecraft due to visualisation purposes, \ie to focus on the sheath and first magnetic structure).
}
\label{fig7}
\end{figure}

Around the time of the CME liftoff several spacecraft at different heliospheric distances were approximately radially aligned with Earth (see Figure \ref{fig7}). We next analyse the \insitu measurements at other heliospheric distances to try to identify the sheath and the structure corresponding to the 2nd step of FD observed by SOHO/EPHIN (hereafter referred to as flux rope, FR). For that purpose we use the MErcury Surface, Space ENvironment, GEochemistry, and Ranging (MESSENGER, MES) magnetometer \citep[MAG,][]{anderson07} and Neutron Spectrometer \citep[NS,][]{goldsten07}, the Venus Express (VEX) magnetometer \citep{zhang06} and the Radiation Assessment Detector \citep[RAD,][]{hassler12}, on board Mars Science Laboratory's (MSL) rover Curiosity \citep{grotzinger12}. The multi-spacecraft \insitu measurements that we compare are shown in Figure \ref{fig7}. At MES magnetic field and cosmic ray counts are available, at VEX only magnetic field and at MSL/RAD only cosmic ray counts. At MES a two-step FD is observed \citep{winslow18}, which helps us to identify the corresponding sheath. It can be seen in Figure \ref{fig7} that the profile of the total magnetic field in the sheath region at MES is remarkably similar to that at Earth. MES was inside Mercury's magnetosphere during the passage of the CME magnetic structure, thus this data was removed. Nevertheless, the end of the rotation is visible after the data gap, as is the recovery phase of the FD, allowing us to set borders to the magnetic structure, presumably corresponding to the FR. At VEX the similarity to MES and Wind data is not obvious and there are no cosmic ray measurements. Nevertheless, we set the borders of the presumed FR based on the rotation of the magnetic field, assuming that the distorted leading part of the FR as observed at Earth is already present at VEX. Finally, at Mars we observe only one decrease, the second step is not observed in FD. Therefore, similarly as \citet{winslow18}, we assume that the entire main phase of the FD at Mars (from onset to the minimum) corresponds to the shock/sheath region. This allows us to mark the shock arrival and set borders for the sheath region. However, we cannot identify the FR.

Based on the identification of sheath and FR at different heliospheric locations as shown in Figure \ref{fig7} we analyse the evolution of the size and magnetic field in the sheath and FR. The size of the sheath was estimated based on the measured sheath duration and average flow speed. The flow speed at MES, VEX and Mars was estimated using the method described by \citet{vrsnak19}, where two speed measurements at two different locations are used to extrapolate the propagation speed at a third location (for that purpose we used the CME initial speed and the flow speed observed at Earth). The FR size was similarly estimated based on the measured FR duration and average flow speed, where the average flow speed was estimated based on the expansion speed. At Earth, the expansion speed was estimated from the linearly decreasing speed profile. We then assumed constant expansion speed and at a given location (Mercury, Venus, Mars) subtracted the expansion speed from the sheath flow speed in order to estimate the average FR flow speed (\ie we assumed the leading edge of the FR has the same speed as the flow in the sheath). In order to add an additional datapoint for the analysis, we also calculated the diameter of the croissant obtained from the GCS fit as a FR size proxy at the Sun, and we used the distance between FR and shock apex obtained from the GCS fit as a proxy of the sheath size at the Sun. The magnetic field strength in the sheath and FR was estimated manually by the observer (M.D.) based on the plot of the time series in the specific region. Although this introduces a certain subjectivity, calculation of \eg average value or simply taking a peak value might yield a misleading result due to changes on fine time scales, which can substantially deviate from the smooth models we use to understand the global structure.

\begin{figure}
\centerline{\includegraphics[width=0.99\textwidth]{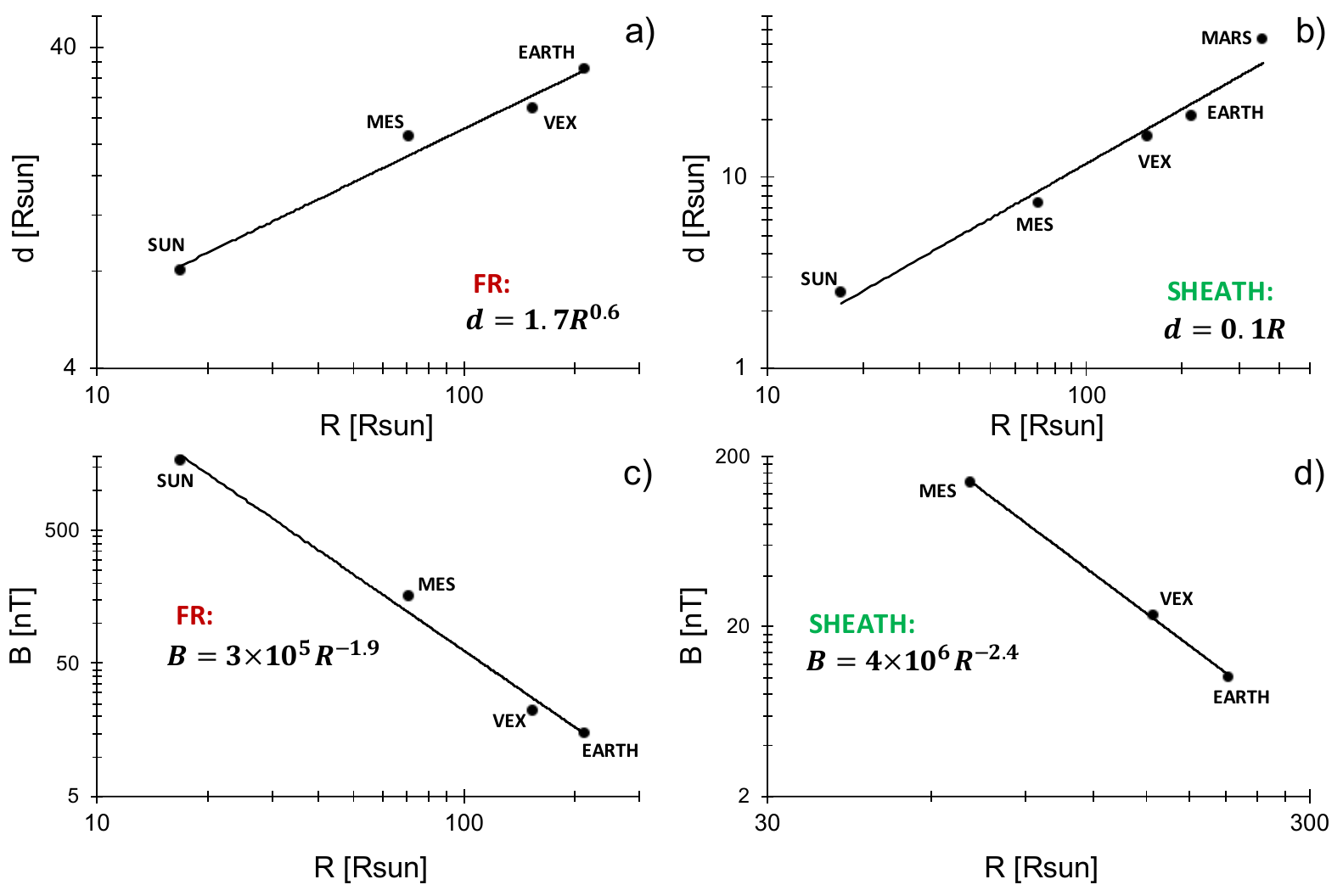}}
\caption{Power-law fits for radially aligned multi-spacecraft data of: a) FR size; b) sheath size; c) FR magnetic field; d) sheath magnetic field (in log-log scale; for details see main text).}
\label{fig8}
\end{figure}

Next, in order to add an additional datapoint for the analysis of the magnetic field, we estimate the magnetic field in the FR at the Sun using the value of the dimming flux (\ie total unsigned magnetic flux involved in the mapped coronal dimming region) calculated at the time of the first eruption observed in SDO (upper plots of Figure \ref{fig5}). We use the dimming flux as a proxy of the magnetic flux contained within the CME magnetic structure \citep{dissauer18a}, similarly as was done by \citet{scolini20}, where the magnetic field inside the structure was estimated using the GCS-reconstructed radius to calculate the cross-section area (we estimate a dimming flux of $\approx4\times10^{21}$Mx). The results are summarized in Table \ref{tab1} and shown in Figure \ref{fig8}, where power-law fits are applied to characterise the evolutionary properties.  It can be seen that the power-law index for the increase of the FR size is at the lower end of the typical observational range \citep{gulisano12}. We also estimate the size expansion factor, \ie the size power-law index, based on the relation presented by \citet{gulisano12}, which applies to non-perturbed magnetic clouds expanding self-similarly ($n_a=\Delta vR/\Delta t\langle v\rangle^2$, where $n_a$ is the size power-law index, $\Delta v$ is the difference between the flow speeds of the leading and trailing edge, $R$ is heliospheric distance, $\Delta t$ is duration and $v$ the flow speed). Thus, the obtained size power-law index is somewhat larger ($n_a=0.8$) than the one obtained in Figure \ref{fig8} ($n_a=0.6$). In addition, we find that the size of the sheath also increases with heliospheric distance and moreover at a faster rate ($n_a=1$) than the size of the FR. This is in agreement with the study of \citet{janvier19}, who found that the ratio of the duration of the magnetic ejecta over that of the sheath in general decreases from MESSENGER to ACE. The magnetic field power-law index for the FR ($n_B=1.9$) is well within the observationally expected range \citep[see][]{gulisano12}, whereas for the sheath it is somewhat larger ($n_B=2.4$).

\begin{table}
\tiny
\caption{CME evolutionary parameters obtained based on multi-spacecraft observation and modelling}
\label{tab1}
\begin{tabular}{lccccc}
\hline
												&	SUN					&	MES				&	VEX				&	EARTH\tabnote{\tbnn}	&	MARS\\
\hline												
heliospheric distance [au/\Rsun]						&	0.08/17				&	0.33/71			&	0.72/155			&	0.99/213				&	1.66/358\\
sheath duration [day/h]								&	--					&	0.1/2.4			&	0.3/7.2			&	0.4/9.6				&	1/24\\
FR duration [day/h]									&	--					&	0.3/7.2			&	0.5/12			&	0.7/16.8				&	--\\
sheath B [nT]										&	--					&	140				&	23				&	10					&	--\\
FR B [nT]											&	1600\tabnote{\tbn}		&	160				&	22				&	15					&	--\\
sheath FDmax [\%]									&	--					&	4				&	--				&	9/4					&	7\\
FR FDmax [\%]										&	--					&	7				&	--				&	2/0					&	0\\
flow speed [\kmps]\tabnote{\tbnnn}						&	--					&	590				&	440				&	420					&	420\\
expansion speed [\kmps]\tabnote{\tbnnnn}					&	--					&	25				&	25				&	25					&	25\\
sheath size [au/\Rsun]								&	--/2.5\tabnote{\tbnnnnn}	&	0.034/7			&	0.076/16			&	0.097/21				&	0.242/52\\
FR size [au/\Rsun]									&	0.04/8				&	0.1/21			&	0.12/26			&	0.16/34				&	--\\
\hline
\textit{ForbMod} results									&						&					&					&						&	\\
\hline
FR FDmax [\%] (ForbMod1\tabnote{\tbnnnnnn} E=1.2GV)		&	--					&	44				&	--				&	3.5					&	0.2\\
FR FDmax [\%] (ForbMod1 $E>E_{\mathrm{cutoff}}$)		&	--					&	16				&	--				&	3.1/0.4				&	0.7\\
FR FDmax [\%] (ForbMod2\tabnote{\tbnnnnnnn} E=1.3GV)	&	--					&	--				&	--				&	2					&	--\\
FR FDmax [\%] (ForbMod2 $E>E_{\mathrm{cutoff}}$)		&	--					&	--				&	--				&	1.7					&	--\\
\hline
\end{tabular}
\end{table}

\subsection{CME evolutionary properties and Forbush decreases}
\label{FD}	

We next study how the FR evolutionary properties correspond to the GCR profiles. Due to the complex nature of the event, which deviates substantially from the generic profile discussed in Section \ref{profile}, we do not analyse the recovery phase. Furthermore, the PDB model for the sheath region in its current form is not suitable for quantitative analysis since it is a steady-state model in which the evolution of the sheath is not taken into account and it is not properly normalised by initial and boundary conditions, so that the allowed input yields an FD result in the range [0\%,100\%]. Therefore, we limit ourselves to the analysis of the FR evolutionary properties and compare  them to the \textit{ForbMod} results. We calculate the expected FD amplitude at SOHO/EPHIN using the observationally constrained FR properties and energy dependence adapted for the SOHO/EPHIN response function, estimated based on the simulations performed by \citet{kuhl15} (for details see Appendix \ref{ephin}). As initial FR input we use the results of the GCS reconstruction ($a_0=4$ \Rsun\, at $R=17$ \Rsun), whereas the initial diffusion coefficient is calculated based on the FR magnetic field measured at Earth ($B=15$ nT) using the procedure explained in Appendix \ref{diff_coeff}. We use the ICME transit time to Earth ($TT=74$h) as the diffusion/expansion time, and observationally obtained expansion indices ($n_B=1.9$, $n_a=0.8$; note that we use the size expansion index as obtained by the \citet{gulisano12} method). The results are presented in Table \ref{tab1}. The calculated FD amplitude ($1.7\%$) is in good agreement with the observed one ($2\%$).

We next test the assumption that in \textit{ForbMod} the energy dependence can be simulated with the monoenergetic model, if one conveniently assumes the rigidity of the particles that the detector is mostly sensitive to, as was previously applied by \citet{rodari18} and \citet{dumbovic19}. This is similar to the concept of effective rigidity to characterise the detector's response \citep{kalugin15,asvestari17,koldobskiy19}, where the effective rigidity is defined as the rigidity level at which the variations in CR flux are the same as the variations integrated over the entire energy range. In the monoenergetic \textit{ForbMod} model we assume that the effective rigidity for a specific event observed in SOHO/EPHIN is defined by the peak of the GCR fractional contribution function, taking into account the SOHO/EPHIN response function. The GCR fractional contribution, calculated based on Equation \ref{eq9} has a rigidity peak at $1.3$GV. The monoenergetic \textit{ForbMod}, using the diffusion coefficient for $1.3$GV particles yields an FD amplitude of $2\%$, which is somewhat larger than the result for the energy-dependent model, but interestingly matches the observations. A possible explanation might lie in the fact that the adiabatic cooling was not included, which, if included, might shift both calculated FD amplitudes to higher values.

We now consider the FD amplitude measured in GCR fluxes with different energy ranges (or cutoffs). In order to calculate FD amplitudes measured with other instruments and at other locations, we calculate the FD amplitude assuming a ``perfect detector", \ie a detector which responds to all energies above the cutoff equally. We note that for many detectors this assumption is invalid, particularly for neutron monitors \citep[\eg][]{clem00,mishev20}. However, as shown in Section \ref{energy}, the FD magnitude calculated for the ``perfect detector" can be taken as a very rough upper limit. The response of some detectors used in the study to GCRs (MES/NS, MSL/RAD) are quite complex and it is not trivial to obtain their response function in a form that can be easily combined with \textit{ForbMod} using the procedure explained in Section \ref{energy}. On the other hand, a cutoff energy of each detector used in the study is known: 0.125 GeV for MES/NS \citep{winslow18}, 0.05 GeV for SOHO/EPHIN \citep{kuhl15}, 0.5 GeV for SoPo \citep{clem00}, and 0.15 GeV for MSL/RAD \citep{guo18c}. The calculated FD amplitude for the SOHO/EPHIN cutoff using the ``perfect detector" approximation is $3.1\%$, which is almost double the value obtained using its response function ($1.7\%$). The calculated FD amplitude at MES distance (\ie using the transit time to MES as diffusion/expansion time) for the MES/NS cutoff using the ``perfect detector" approximation is $16\%$, which is larger than the observed value by about a factor of 2 - similarly as is for SOHO/EPHIN. The calculations at Earth and Mars, using the SoPo and MSL/RAD cutoffs respectively, in the ``perfect detector" approximation, yield FD amplitudes $<1\%$, which is basically within the observational error (given the daily GCR variations).

\begin{figure}
\centerline{\includegraphics[width=0.6\textwidth]{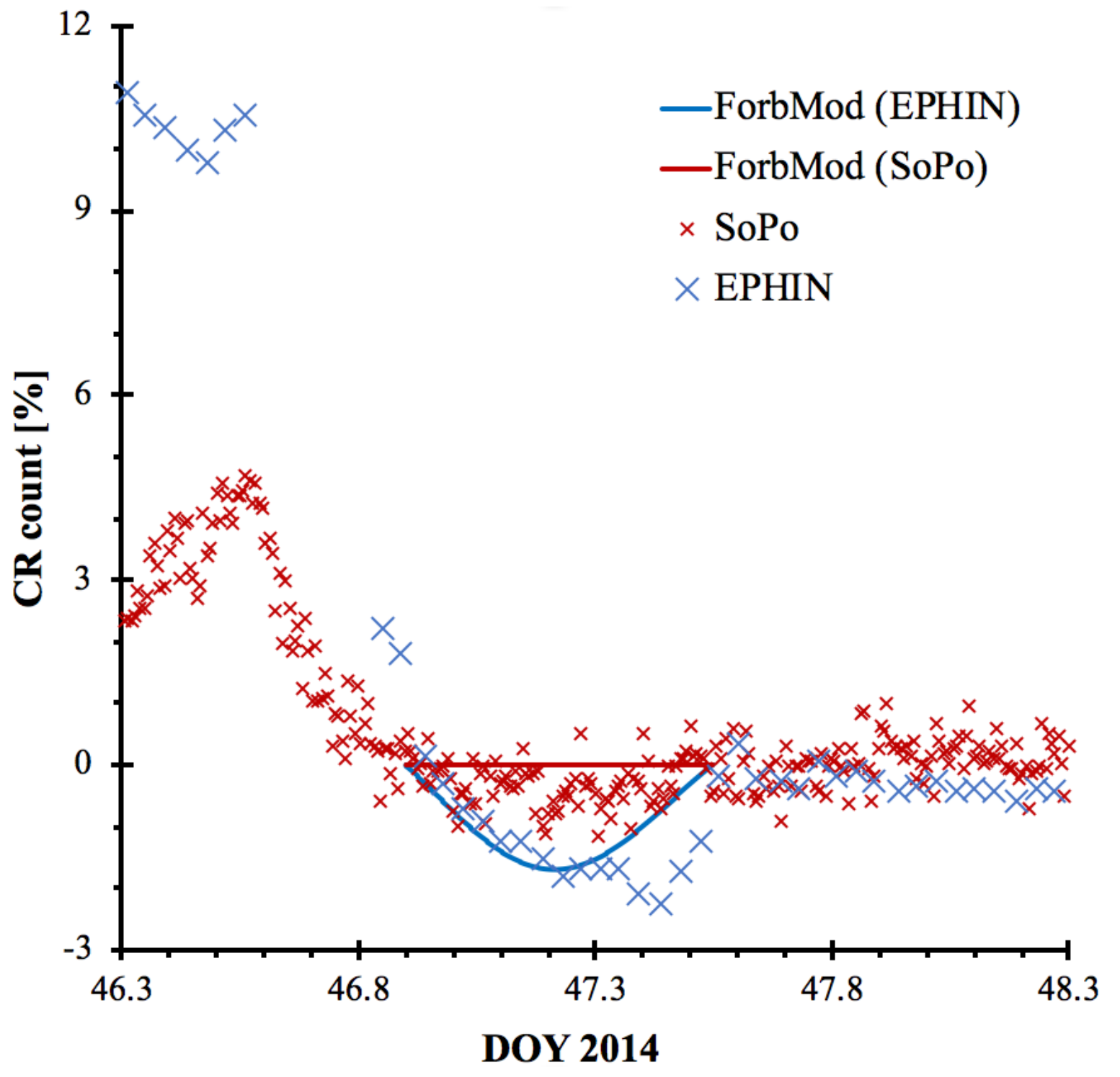}}
\caption{The energy-dependent \textit{ForbMod} results for SoPo (red line) and SOHO/EPHIN (blue line) against observations with SoPo (red crosses) and SOHO/EPHIN (blue crosses) for the FR region of the 2014 February 15 FD. The relative counts have been normalised to the CR count at the start of the FR. FD profile is converted from radial dependence into time dependence based on the ICME speed.
}
\label{fig9}
\end{figure}

We note that the ``perfect detector" approximation by default increases the FD amplitude, as it assumes the same contribution of low and high-energy particles, where the fractional contribution to the total FD amplitude is substantially lower for high-energy particles. Including the response function in the calculation, which reflects lower sensitivity to low-energy particles, would thus further decrease the FD amplitude. Therefore, \textit{ForbMod} calculations for SoPo and Mars are in agreement with observations, since no notable FR-related FD amplitude was measured. Applying the monoenergetic approximation to the ``perfect detector" yields quite unrealistic results, as can be seen in Table \ref{tab1}, related to the fact that the peak of the GCR fractional contribution, is shifted to lower rigidity. We stress that the uncertainties of the observational methodology used are large. Nevertheless, we point out that \textit{ForbMod} gives a quite reasonable agreement with measurements at two locations at Earth, and at Mars. In Figure \ref{fig9} we show the comparison of the SoPo and SOHO/EPHIN measurements with the corresponding energy-dependent \textit{ForbMod} FD profiles. We note that similar comparison cannot be performed for Mars, because we do not know the duration of the FR.

\section{Conclusion and Summary}
\label{conclusion}	

We analyse whether, and how, FDs reflect the evolutionary properties of CMEs. In order to understand the mechanisms which govern the formation of the depression, we analyse separately the interplanetary structures that are the sources of FDs and how they influence the GCRs. We first produce a generic profile of the text-book example two-step FD using three different models for three different regions (sheath, flux rope and post-CME FD recovery region), to see how well it matches the observed FD profiles. We combine two analytical models for the FD main phase (from onset to the minimum), the propagative diffusive barrier (PDB) model for the sheath region and the diffusion-expansion Forbush decrease (\textit{ForbMod}) model for the CME magnetic structure. The recovery phase of the FD is modelled combining the \textit{ForbMod} model for the CME magnetic structure and an exponential recovery caused by the shadow effect of the propagating shock. We find that the modelled generic FD profile describes well the observed two-step decreases, and is, in addition, also able to explain why two-step decreases are not very frequently observed. Our modelling efforts show that the transition points between different regions can be smeared out, making the whole FD appear as a homogeneous phenomenon. 

We next adapt the analytical models describing the FD main phase for energy dependence, in order to compare the modelled results quantitatively with measurements. This is achieved by allowing the diffusion coefficient to be a function of rigidity as well as time. It is shown that with this new adaptation, the modelled FDs are rigidity dependent in agreement with the observations, \ie the depression will be larger for lower energy particles. Moreover, it is shown that the contribution to the total FD from particles of different energies comes from a quite constrained energy-range of particles, due to folding of the FD energy dependence and the GCR spectrum. The distribution of the GCR fractional contribution to the total FD shows bell-curve behaviour with a distinct peak, therefore, allowing the FD calculation to be made mono-energetically, that is, by making the approximation that the main contribution to the total FD comes from particles of specific rigidities. This assumption was used for \textit{ForbMod} analysis in previous studies with reasonable arguments. However, here we tested it for the first time. Comparison of the \textit{ForbMod} results with full energy integration and with monoenergetic approximation for SOHO/EPHIN shows 15\% difference between calculated FD amplitudes. This indicates that using the effective rigidity corresponding to the peak of the GCR fractional contribution can only be taken as a very rough approximation in calculating the total FD amplitude.

Finally, we perform an in-depth study of the multi-spacecraft event to analyse and characterise the CME evolution and apply CME magnetic structure observational characteristics to simulate the corresponding FD. We note that one of the key factors in understanding the inner structure of the CME/ICME event was the FD substructure observed by SOHO/EPHIN. Our modelling results show reasonable agreement with measurements near Earth, at Earth, and at Mars, indicating that FD models not only offer an opportunity to understand the variability of FDs detected in the heliosphere, but also to gain insight into the CME evolution.

 \appendix
 \section{Diffusion coefficient}
 \label{diff_coeff}

\begin{figure}
\centerline{\includegraphics[width=0.99\textwidth]{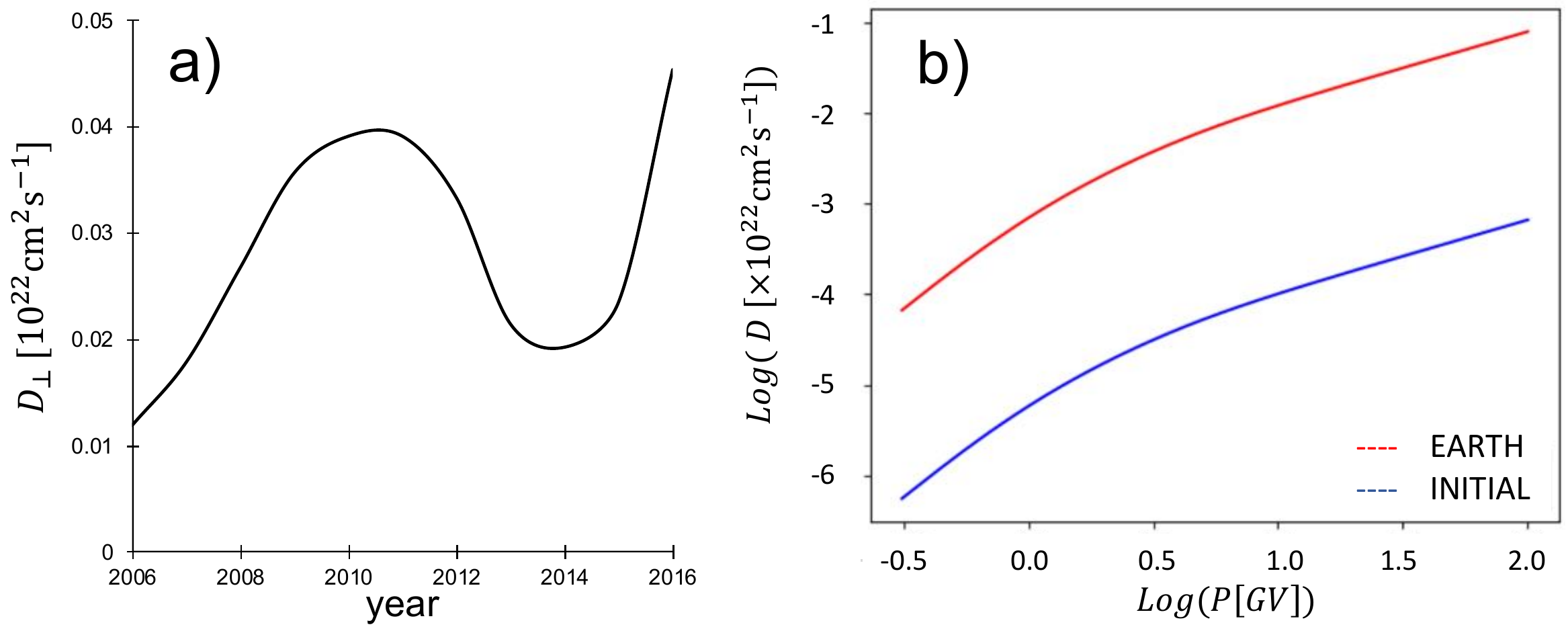}}
\caption{a) perpendicular diffusion coefficient at Earth calculated based on \citet{potgieter14} for 2006--2009 and \citet{corti19} for 2011--2016 (extrapolated to 2010) for magnetic field strength $B=5$ nT and rigidity $P=1$ GV. b) rigidity dependence of the diffusion coefficient in 2014 at Earth and at $R_0=15$\Rsun}
\label{figA}
\end{figure}

In order to introduce energy dependence we allow the diffusion coefficient, $D$, to be a function of rigidity as well as time, which can be expressed through an empirical formula as used in numerical models fitted to GCR measurements, as given by \eg \citet{potgieter13}:

\begin{equation}
 D_E(P)=0.02\cdot10^{22}\cdot k_{\mathrm{||},0}\cdot\beta\frac{P^a}{B}\Bigg[\frac{P^c+(P_k)^c}{1+(P_k)^c}\Bigg]^{(\frac{b-a}{c})}\,,
\label{eqA}
\end{equation}

\noindent where $D_E$ is given in units $\mathrm{cm}^2\mathrm{s}^{-1}$, $P$ is rigidity in units GV, $B$ is the magnetic field in units nT, and $k_{\mathrm{||},0}$, $a$, $b$, $c$ and $P_k$ are parameters obtained empirically from the observation of the GCR spectrum using instruments such as \textit{Payload for Antimatter Matter Exploration and Light-nuclei Astrophysics} \citep[PAMELA,][]{adriani11} on board the Russian Resurs-DK1 satellite as by \citep{potgieter14} or the \textit{Alpha Magnetic Spectrometer} \citep[AMS-02,][]{aguilar13} experiment on board the International Space Station as by \eg \citet{corti19}. We note that the parameters and the dependence in Equation \ref{eqA} is slightly different for these two studies involving PAMELA and AMS. It should be noted that \citet{potgieter14} studied the period of solar minimum 2006--2009, whereas \citet{corti19} studied the period around and after the solar maximum, 2011--2017. It is reasonable to assume that the perpendicular diffusion coefficient changes periodically with time (\ie solar activity) not only due to the change of the IMF strength, but also the time-varying orientation and complexity of the IMF, closely related to the time-varying speeds and densities of the solar wind and reflected in the change of the parameters in Equation \ref{eqA}. In Figure \ref{figA}a we combine the results for the perpendicular diffusion coefficient at Earth, $D_E$, calculated based on \citet{potgieter14} and \citet{corti19} for magnetic field strength $B=5$ nT and rigidity $P=1$ GV, where it can be seen that the resulting $D_E$ varies periodically in rough anti-correlation to the solar activity indicating that the two empirical formulas corresponding to these two different time-periods can be combined. These two studies therefore provide a calculating frame for the energy-dependent diffusion coefficients. In Figure \ref{figA}b we show the rigidity dependence of the diffusion coefficient at Earth in 2014 calculated in this way, as well as the rigidity dependece of the initial diffusion coefficient, estimated at $R_0=15\mathrm{R_{SUN}}$ based on $D_0=D(R(t)/R_0)^{-n_B}$ assuming a magnetic field expansion factor $n_B=1.8$.

 \section{The GCR spectral intensity}
 \label{force_field}

The ``force--field" approximation is used to describe the long-term GCR modulation and is typically valid for quiet-time periods. However, \citet{usoskin15} have shown that the same approximation can be used to describe the GCR spectrum during an FD. In this approximation all GCR modulation mechanisms are gathered into a single parameter called the modulation potential, $\Phi$, which influences the unmodulated local interstellar spectrum $J_{\mathrm{LIS}}$ to yield the time--dependent differential energy spectrum of GCRs as observed near Earth:

\begin{equation}
J(E,\Phi)=J_{\mathrm{LIS}}(E+\Phi)\frac{(E)(E+2m_0)}{(E+\Phi)(E+\Phi+2m_0)}\,,
\label{eqB1}
\end{equation}

\begin{equation}
J_{\mathrm{LIS}}(E)=\frac{1.9\times10^4\cdot P(E)^{-2.78}}{1+0.4866P(E)^{-2.51}}\,,
\label{eqB2}
\end{equation}

\noindent where we assume all GCRs are protons, $E$ is their kinetic energy, $m_0$ their rest mass, $P(E)=\sqrt{E(E+2m_0)}$ the rigidity and $\Phi$ is the modulation potential which is time-dependent and can be obtained empirically based on GCR measurements \citep{usoskin11,usoskin17}. However, it was shown by \citet{gieseler17} that it is not sufficient to describe GCR intensities at Earth by only one rigidity-independent parameter $\Phi$, as it also depends on the energy range of interest and there are severe limitations at lower energies. Therefore, we use a modified force field approach by \citet{gieseler17} in which the rigidity-dependent modulation parameter is given by:

\begin{equation}
	\Phi(P) =
	\begin{cases}
	\frac{\Phi_{\mathrm{USO11}}-\Phi_{\mathrm{PP}}}{P_{\mathrm{USO11}}-P_{\mathrm{PP}}}\cdot(P-P_{\mathrm{PP}})+\Phi_{\mathrm{PP}}, P<P_{\mathrm{USO11}}\\
	\Phi_{\mathrm{USO11}}, P\ge P_{\mathrm{USO11}}
	\end{cases}
\label{eqB3}
\end{equation}

\noindent where $\Phi_{\mathrm{USO11}}$ is the solar modulation potential obtained for neutron monitors empirically by \citet{usoskin11}, $\Phi_{\mathrm{PP}}$ is the solar modulation potential derived from the 1.28 GV proton proxies IMP-8 helium and ACE/CRIS carbon by \citet{gieseler17}, and $P_{\mathrm{USO11}}=13.83\pm4.39$GV and $P_{\mathrm{PP}}=1.28\pm0.01$GV are the corresponding mean rigidities. Equations \ref{eqB1} --\ref{eqB3} therefore provide a scheme to calculate GCR spectrum for a given event. We note that for the event presented in Section \ref{multi-spacecraft} the uncorrected and corrected solar modulation potentials are 0.681 and 0.97 GV, respectively \citep{gieseler17}.

 \section{The SOHO/EPHIN response function}
 \label{ephin}

\begin{figure}
\centerline{\includegraphics[width=0.8\textwidth]{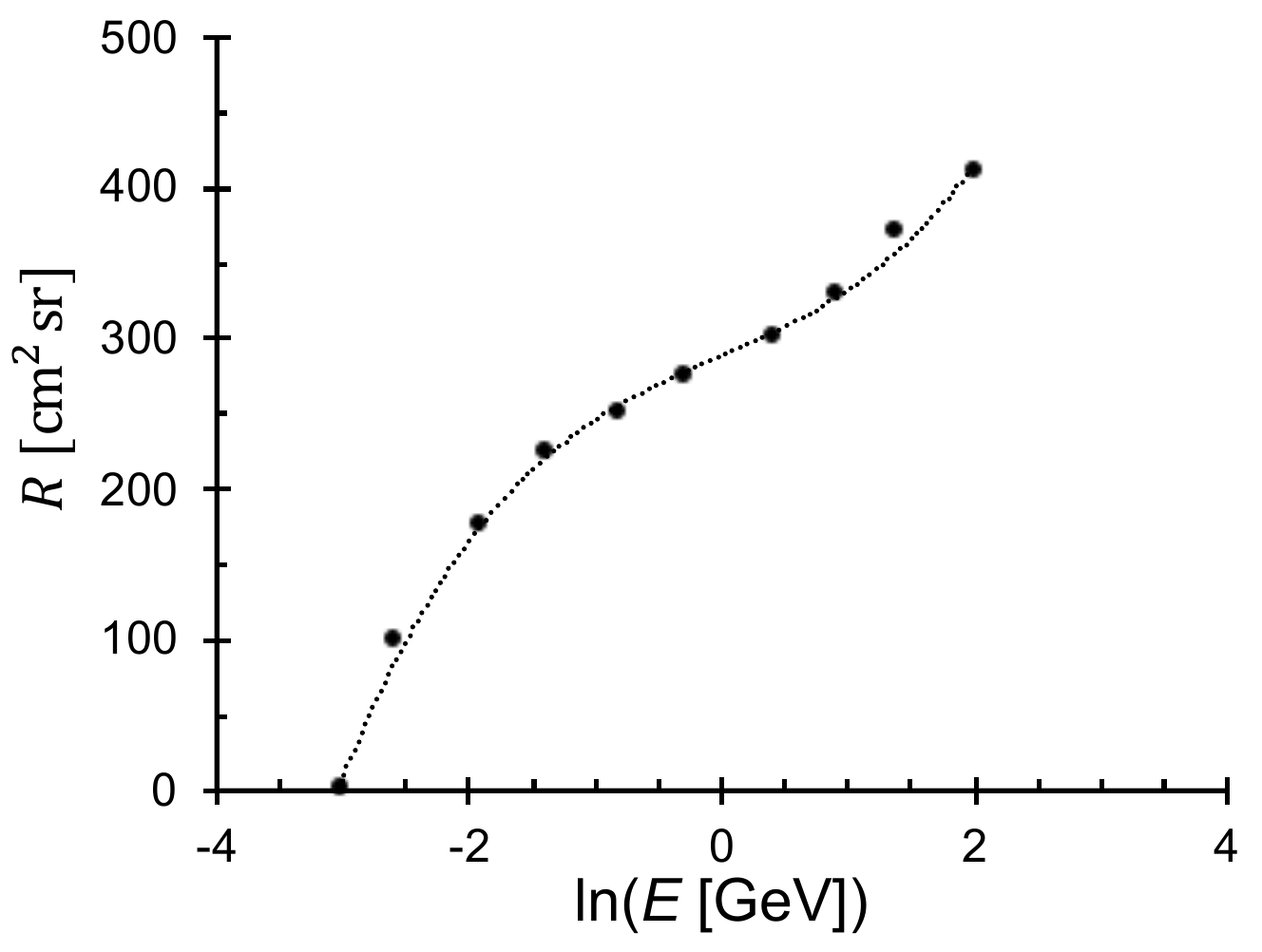}}
\caption{GEANT 4 Monte Carlo simulation of the response function of the single detector F of the SOHO/EPHIN instrument (black dots) and a fitting function approximating its analytical form (dotted line).}
\label{figC}
\end{figure}

In order to obtain an analytical form of the response function of the single detector F of the SOHO/EPHIN instrument, we use the GEANT 4 Monte Carlo simulation of the instrument performed by \opencite{kuhl15} for the omnidirectional isotropic flux of protons, given in Figure 1 of \citet{kuhl15}. The data points representing the GEANT 4 simulation are presented in Figure \ref{figC}, where a fitting function is applied in order to derive an approximate analytical form of the functional dependency:

\begin{equation}
 R = 6.4\ln^3(E)-0.42\ln^2(E)+36.8\ln(E)+288.8
\label{eqC1}
\end{equation} 

%
\begin{acks}
The research leading to these results has received funding from the European Unions Horizon 2020 research and innovation programme under the Marie Sklodowska-Curie grant agreement No 745782 (ForbMod). B.V. and M.D. acknowledge a support by the Croatian Science Foundation under the project 7549 (MSOC). J.G. acknowledge the Strategic Priority Program of the Chinese Academy of Sciences (Grant No. XDB41000000 and XDA15017300)  and the CNSA pre-research Project on Civil Aerospace Technologies (Grant No. D020104). B. H. acknowledges the discussions from the HEROIC team at the International Space Science Institute. K.D. and A.M.V. acknowledge funding by the Austrian Space Applications Programme of the Austrian Research Promotion Agency FFG: projects ASAP-11 4900217 and ASAP-14 865972. F.C. acknowledges the financial support by MINECO-FPI-2016 predoctoral grant with FSE, and its project FEDER/MCIU-AEEI/Proyecto ESP2017-88436-R. C.M. and T.A. thank the Austrian Science Fund (FWF): P31659-N27, P31521-N27, P31265-N27. We acknowledge the NMDB database (http://www.nmdb.eu) founded under the European Unions FP7 programme (contract No. 213007), and the PIs of individual neutron monitors for providing SoPo data. MESSENGER and MSL data are available on the Planetary Data System (https://pds.jpl.nasa.gov). SOHO/EPHIN is supported by the Ministry of Economics via DLR grant 50OG1702. We thank Ewan Dickson, PhD for improving the readability of the paper. Finally, we thank the anonymous reviewer whose thorough revision and insightful comments significantly improved the quality of the paper.
\end{acks}

\section*{Disclosure of Potential Conflicts of Interest}
The authors declare that they have no conflicts of interest.

 \bibliographystyle{spr-mp-sola}
 \bibliography{REFs}  

\end{article} 
\end{document}